\documentclass[aps,pra,twocolumn,showpacs,superscriptaddress,longbibliography]{revtex4-1}
\usepackage{graphicx}

\usepackage{amsmath}

\usepackage{amssymb}

\usepackage{epstopdf}

\usepackage{amsfonts}

\usepackage{dcolumn}

\usepackage{multirow}

\usepackage[utf8]{inputenc}
\usepackage{esint}
\usepackage{float}

\makeatletter
\@ifundefined{textcolor}{}
{
 \definecolor{BLACK}{gray}{0}
 \definecolor{WHITE}{gray}{1}
 \definecolor{RED}{rgb}{1,0,0}
 \definecolor{GREEN}{rgb}{0,1,0}
 \definecolor{BLUE}{rgb}{0,0,1}
 \definecolor{CYAN}{cmyk}{1,0,0,0}
 \definecolor{MAGENTA}{cmyk}{0,1,0,0}
 \definecolor{YELLOW}{cmyk}{0,0,1,0}
}

\begin{document}

\title{Spinor Bose-Einstein 
Condensate
Interferometer within the Undepleted Pump Approximation: 
Role of the Initial State}

\author{Jianwen Jie}
\email{Jianwen.Jie1990@gmail.com}
\affiliation{Homer L. Dodge Department of
Physics and Astronomy, The University of Oklahoma, 
440 W. Brooks Street, Norman, Oklahoma 73019,
USA}
\author{Q. Guan}
\email{gqz0001@gmail.com}
\affiliation{Homer L. Dodge Department of
Physics and Astronomy, The University of Oklahoma, 
440 W. Brooks Street, Norman, Oklahoma 73019,
USA}
\author{D. Blume}
\email{doerte.blume-1@ou.edu}
\affiliation{Homer L. Dodge Department of
Physics and Astronomy, The University of Oklahoma, 
440 W. Brooks Street, Norman, Oklahoma 73019,
USA}
\date{05/04/2019}
\begin{abstract}
Most interferometers operate with photons or dilute,
non-condensed cold atom clouds in which collisions are 
strongly suppressed.
Spinor Bose-Einstein condensates (BECs)
provide an alternative  
route toward realizing 
three-mode interferometers; in this realization, spin-changing
collisions provide a resource that generates mode entanglement.
Working in the regime where the pump mode,
i.e., the $m=0$ hyperfine state, has a much larger
population than the side or probe modes 
($m= \pm 1$ hyperfine states),
$f=1$ spinor BECs approximate SU(1,1) 
interferometers.
We derive analytical expressions 
within the undepleted pump approximation
for the phase sensitivity
of such an SU(1,1) interferometer
for two classes of initial states: pure Fock states and coherent
spin states.
The interferometer performance is analyzed for initial states without
seeding, with single-sided seeding, and with double-sided
seeding. 
The validity regime of the undepleted pump approximation
is assessed by performing quantum calculations for the 
full spin Hamiltonian.
Our analytical results and the associated dynamics are
expected to guide experiments as well as numerical
studies that explore regimes where the undepleted pump
approximation makes quantitatively or qualitatively incorrect
predictions.
\end{abstract}

\maketitle
\section{Introduction}

Quantum enhanced measurement protocols or
quantum metrology refer to improving the precision measurement
of a physical parameter or physical parameters
using quantum protocols~\cite{RMP2018:Smerzi,RMP2018_HL:Braun}. 
Nowadays, quantum metrology is a powerful workhorse across physics, 
including areas as diverse as gravitational wave 
detection~\cite{RevModPhys.52.341,PRD1981_SQL:Caves,Pitkin2011}; 
sensing applications~\cite{RevModPhys.89.035002} such as 
magnetometry~\cite{PhysRevLett.104.133601,PhysRevX.5.031010}, 
gravitometry~\cite{Freier_2016}, and 
electric field determinations~\cite{Fan_2015}; 
optical communication~\cite{6999929}; and image 
reconstruction~\cite{5355500,NaturePhotonicsvolume4pages227to230_2010}.

A classical approach for improving the
estimation would repeat the measurement on
$N$ identical but independent systems or uncorrelated particles.
For single parameter estimations,
such an approach leads to a $1/\sqrt{N}$ scaling,
which is typically referred to as standard quantum 
limit~\cite{PRD1981_SQL:Caves} or shot noise 
limit~\cite{PRA1986_SU11:Yurke,PRL1987_SNL:Xiao}. 
Beyond (i.e., better than) the standard quantum limit performance 
can be achieved by taking advantage of quantum resources.
Caves pointed out in 1981~\cite{PRD1981_SQL:Caves} 
that squeezed states 
can improve the performance to a $1/N$ scaling.
Motivated by
the heuristic phase-particle number Heisenberg
uncertainty relation 
$\Delta \theta \Delta N\geq1$,
Holland and Burnett~\cite{PRL1993_HL:Burnett} referred to the
$1/N$ performance as
``Heisenberg limit''.
Unfortunately, unique definitions of the standard 
quantum limit and the Heisenberg limit are not available~\cite{RMP2018_HL:Braun}. 
Quite generally, to specify these limits, 
the
classical resources need to be defined
and the
improvement of the parameter estimation
due to the additional quantum resources
needs to be quantified.

Assuming a generic set-up that consists of three components---(i) input, 
(ii) ``actual device'', and (iii) measurement and 
parameter estimation---one can attempt to improve the performance by optimizing either of
the three components listed above. The present work focuses on quantifying the performance 
of a paradigmatic device, namely an interferometer, using 
established formulations for the parameter estimation: the 
phase sensitivities $\Delta \theta_{\text{QCR}}$ and $\Delta \theta_{\text{ep}}$ that are, respectively, derived from the quantum Cramer-Rao bound~\cite{PRL1994_QFI:Braunstein,PRL2009_Hl:Pezz,PRA2009_HL:Boixo} and error propagation~\cite{RMP2018:Smerzi}. 

The interferometer considered is an SU(1,1) interferometer based on a $f=1$ spinor Bose-Einstein condensate (BEC)~\cite{PR2012_Spinor_BEC:Ueda,RMP2013_Spinor_BEC:Ueda} with three internal hyperfine components, namely the hyperfine states with projection
quantum numbers $m=+1$, $m=0$, and $m=-1$
($f$ denotes the total spin angular momentum of the atom). 
An SU(1,1) interferometer can be constructed by replacing the passive beam splitters 
in a Mach-Zehnder interferometer by active non-linear parametric amplifiers,
which can generate 
quantum correlations and entanglement~\cite{PRA1986_SU11:Yurke}.

Our study is motivated by the quest to get a handle on the role played by correlations and entanglement
of the initial state
and of the state
during the amplification step. 
Given a device and parameter estimation scheme,
how does the absolute performance depend on the initial state?
Given a certain class of initial states, what are the device  
parameters that yield the best absolute phase sensitivity?
Besides providing general insight and being important for studies 
in the regime where the
undepleted pump approximation holds,
our results are expected to provide guidance for
spinor BEC based interferometer studies 
that operate outside the SU(1,1) regime.

Treating the spinor BEC in the single-mode approximation~\cite{PRA2002_SMA:Yi} and further working 
in the 
undepleted pump approximation~\cite{PRL2015_SU11:Gabbrielli}, we report explicit analytic expressions for the phase sensitivity and a number of auxiliary observables for two classes of initial states: pure Fock states and coherent spin states. 
We consider the situations where all atoms are in the $m=0$ state and the side modes are 
empty (``vacuum state''~\cite{PRL2016_SU11:Linnemann}), 
the majority of atoms is in the $m=0$ state and either the 
$m=+1$ mode or $m=-1$ mode has a small population (single-sided seeding), 
and the majority of atoms is in the $m=0$ state and the 
$m=+1$ and $m=-1$ modes both have small populations (double-sided seeding). 
For selected observables, we present analytic expressions for an arbitrary
pure initial state.
Physical interpretations of the analytical expressions
are presented. The validity regime of the undepleted pump approximation 
is assessed by 
simulating the entire interferometer 
sequence for the full spin Hamiltonian numerically. 

The remainder of this paper is organized as follows.
Sections~\ref{sec_spinham} and \ref{sec_IIB} 
introduce the spin Hamiltonian that underlies this work
and the basic operating principle of a spinor
BEC based interferometer, respectively.
The parameter estimation
procedures considered in this work 
are introduced 
in
Sec.~\ref{sec_param}
and the undepleted pump approximation 
is introduced in Sec.~\ref{sec_upa}.
The 
equations of motion within this approximation and their
solutions are
introduced in
Secs.~\ref{sec_eom} and \ref{sec_casei}
and explicit analytical results for
the side mode population and the corresponding fluctuation
are presented in Sec.~\ref{sec_sidemodepopulation}.
Section~\ref{sec_results}
analyzes our analytical results.
Two classes of initial states
are considered:
pure Fock states (see Sec.~\ref{sec_results_pfs})
and coherent spin states
(see Sec.~\ref{sec_results_css}).
Finally, Sec.~\ref{sec_conclusion}
presents a conclusion.

\section{Problem definition and theoretical background}
\label{sec_theory}

\subsection{Full spin Hamiltonian}
\label{sec_spinham}
Our description accounts for the three hyperfine states
of an $f=1$ spinor BEC consisting of $N$
atoms within the single-mode approximation,
which assumes that the spatial degrees of freedom are integrated
out~\cite{PRL1998_SMD:Law,PRA2002_SMA:Yi}. 
As a result, the dynamics is governed by the spin Hamiltonian
$\hat{H}_{\text{spin}}$, which treats each
atom as a structureless spin-1 ``object'' that undergoes 
two-body $s$-wave collisions~\cite{PRL1998_SMD:Law},
\begin{eqnarray}
\label{eq_hspin}
\hat{H}_{\text{spin}}(c,q)&=&\frac{c}{N} \left(\hat{a}^{\dagger}_{+1}\hat{a}^{\dagger}_{-1}\hat{a}_{0}\hat{a}_{0}+\mbox{h.c.}\right)\nonumber\\
&&+\frac{c}{N} \left(\hat{N}_0-\frac{1}{2}\right)\left(\hat{N}_{+1}+\hat{N}_{-1}\right)\nonumber\\
&&+q\left(\hat{N}_{+1}+\hat{N}_{-1}\right).
\end{eqnarray}
The first term on the right hand side of Eq.~(\ref{eq_hspin})
describes spin-changing collisions, also referred
to as spin-mixing dynamics; this term 
is identical to the four-wave mixing term 
in nonlinear quantum optics~\cite{PRA1986_SU11:Yurke}.
The second term on the right hand side of Eq.~(\ref{eq_hspin})
corresponds
to the collisional shift and the third term is a single-particle shift. 
For reasons that will become clear in Sec.~\ref{sec_IIB},
we refer to this term of the Hamiltonian
as linear phase shifter (LPS) Hamiltonian $\hat{H}_{\text{LPS}}$,
\begin{eqnarray}
\hat{H}_{\text{LPS}}(q)=q\left(\hat{N}_{+1}+\hat{N}_{-1}\right).
\end{eqnarray}
The operators $\hat{a}_{m}^{\dagger}$ and
$\hat{a}_{m}$ 
satisfy the bosonic commutation relation 
$\left[\hat{a}_{m},\hat{a}_{m}^{\dagger}\right]=1$
($\hat{a}_{m}^{\dagger}$ 
creates and $\hat{a}_{m}$ destroys an atom in hyperfine state
$|f=1,m \rangle$) and 
the atom number operator $\hat{N}_{m}$
is defined through 
$\hat{N}_{m}=\hat{a}^{\dag}_{m}\hat{a}_{m}$.
The coefficient $q$
contains a ``Zeeman contribution'' $q_{{\rm{B}}}$ from an
external
magnetic field and a contribution
$q_{{\rm{MW}}}$ from a microwave field,
$q=q_{{\rm{B}}}+q_{{\rm{MW}}}$~\cite{PRA2006_MW_dressing:Gerbier,PRA2014_MW_dressing:Zhao}.
The strength $c$, 
\begin{eqnarray}
\label{eq_cbar_to_c}
c=\overline{c} \, \overline{n},
\end{eqnarray} 
of the collision terms is determined by the
mean spatial density $\overline{n}$ and the coefficient $\overline{c}$, 
which is proportional to
the difference between the scattering lengths $a_{F}$ for 
two atoms with total spin angular momentum $F=0$ and $F=2$,
\begin{eqnarray}
\overline{c}=\frac{2\pi \hbar^2}{\mu}\frac{a_2-a_0}{3}.\label{c2value}
\end{eqnarray}
Here, $\mu$ is the reduced two-body mass.
In typical $^{23}$Na and $^{87}$Rb 
BEC experiments, $|c/h|$ is of the order of $20$~Hz 
($\overline{c}$ and $c$ are both positive for $^{23}$Na and
both negative for
$^{87}$Rb) and
$q/h$ can be tuned from negative values to zero to values
much larger than $|c/h|$~\cite{PRA2006_MW_dressing:Gerbier,PRA2014_MW_dressing:Zhao}.

\subsection{Spinor BEC interferometer}
\label{sec_IIB}
\begin{figure}
  \centering
  \vspace*{0.in}
\includegraphics[angle=90,width=0.45\textwidth]{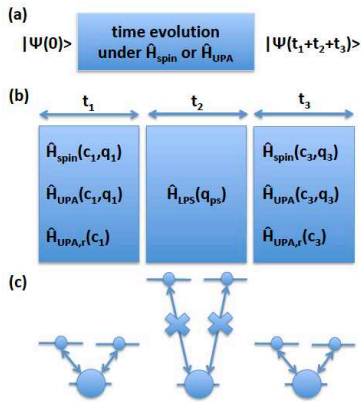}
\vspace*{0.0in}
\caption{(Color online)
Schematic of interferometer.
(a) Basic layout.
The initial state $| \Psi(0) \rangle$
gets fed into the interferometer at time $t=0$.
At time $t=t_1+t_2+t_3$, measurements are made
on $|\Psi(t_1+t_2+t_3)\rangle$.
The present paper investigates how the
phase sensitivity $\Delta \theta_{\text{ep}}$, obtained from
$|\Psi(t_1+t_2+t_3)\rangle$,
depends on the initial state $| \Psi(0) \rangle$
and how the phase sensitivity can be minimized
for a given $|\Psi(0)\rangle$ by tweaking the parameters of
the Hamiltonian that governs the time evolution.
(b) Break-down of the three-step interferometer sequence.
During step 1., the dynamics is
governed by $\hat{H}_{\text{spin}}(c_1,q_1)$,
$\hat{H}_{\text{UPA}}(c_1,q_1)$, or
$\hat{H}_{\text{UPA,r}}(c_1)$.
During step 2., the dynamics is governed by
the linear phase shifter Hamiltonian
$\hat{H}_{\text{LPS}}(q_{\text{ps}})$.
During step 3., the dynamics is
governed by $\hat{H}_{\text{spin}}(c_3,q_3)$,
$\hat{H}_{\text{UPA}}(c_3,q_3)$, or
$\hat{H}_{\text{UPA,r}}(c_3)$.
(c) 
The horizontal lines represent, from left to right, the
hyperfine states $m=+1$, $m=0$, and $m=-1$.
The population of the $m=0$ mode is much larger than that
of the $m= \pm 1$ side modes.
Spin-changing collisions play an important role
during steps 1. and 3. of the interferometer
sequence but are 
turned off during step 2..
   }\label{fig_schematic}
\end{figure}

The three-mode spinor BEC interferometer 
takes an initial state $| \Psi(0) \rangle$, time
evolves it under the spin Hamiltonian $\hat{H}_{\text{spin}}$,
and then performs a measurement or
measurements that form the basis for
determining
the phase sensitivity
[see Fig.~\ref{fig_schematic}(a)].
In our work, $| \Psi(0) \rangle$ is
a
pure state; more generally, 
one could consider a mixed initial state $\hat{\rho}(0)$.
The time evolution is, as shown
in Fig.~\ref{fig_schematic}(b), divided into three time intervals
of lengths $t_1$, $t_2$, and $t_3$.
\begin{enumerate}
\item The first time sequence ($t=0$ to $t=t_1$), which is referred to
as
``state preparation'', applies $ \hat{H}_{\text{spin}}(c_1,q_1)$ to
the initial state $|\Psi(0) \rangle$,
\begin{eqnarray}
|\Psi(t_1)\rangle=
e^{-\imath \hat{H}_{\text{spin}}(c_1,q_1)t_{1}/\hbar}|\Psi(0)\rangle.
\label{interferometer_1}
\end{eqnarray}
\item 
The second, ``phase encoding'' time sequence ($t=t_1$ to
$t=t_1+t_2$)
imprints the relative phase $\theta=2q_{\text{ps}}t_2/\hbar$ 
by applying the linear phase shifter 
Hamiltonian $\hat{H}_{\text{LPS}}(q_{\text{ps}})$,
which is characterized by the generator
$(\hat{N}_{+1} + \hat{N}_{-1})/2$~\cite{PRA2009_HL:Boixo},
\begin{eqnarray}
|\Psi(t_1+t_2)\rangle=
e^{-\imath \hat{H}_{\text{LPS}}(q_{\text{ps}})t_2/\hbar}
|\Psi(t_1)\rangle.\label{interferometer_2}
\end{eqnarray}
If an appropriate Fano-Feshbach resonance~\cite{RMP2010_FR:Chin} exists,
the linear phase shifter Hamiltonian can be realized by 
tuning $c$ to zero.
In the absence of
Fano-Feshbach resonances, the linear phase shifter Hamiltonian
can be realized approximately by
operating in the regime where  $|q| \gg |c|$.
\item 
The third time sequence ($t=t_1+t_2$ to $t=t_1+t_2+t_3$),
which is referred to
as
``read out'', applies the
spin-mixing Hamiltonian $\hat{H}_{\text{spin}}(c_3,q_3)$,
\begin{eqnarray}
~~~~~~~|\Psi(t_1+t_2+t_3)\rangle=e^{- \imath \hat{H}_{\text{spin}}(c_3,q_3)t_{3}/\hbar}
|\Psi(t_1+t_2)\rangle.~~~~\label{interferometer_3}
\end{eqnarray}
\end{enumerate}

The mean number $N_s(t)$ of atoms in the
$m=+1$ and $-1$ side modes
and
the corresponding quantum fluctuation $\Delta N_s(t)$
play an important role in analyzing the
interferometer performance,
which is
quantified by the phase sensitivity (see Sec.~\ref{sec_param}
for details). 
We define
\begin{eqnarray}
\hat{N}_{s}=\hat{N}_{+1}+\hat{N}_{-1},
\end{eqnarray}
\begin{eqnarray}
N_s(t)=\langle\hat{N}_{s} \rangle
= \langle \Psi(t) | \hat{N}_s | \Psi(t) \rangle,
\label{def_Ns}
\end{eqnarray}
and
\begin{eqnarray}
\Delta N_{s}(t)=
\sqrt{\langle (\hat{N}_{s})^{2}\rangle-\langle\hat{N}_{s}\rangle^{2}}.\label{def_delta_Ns}
\end{eqnarray}
The quantities $N_m(t)$ and their quantum fluctuations
are defined analogously.

\subsection{Quantifying the interferometer performance}
\label{sec_param}
To quantify the interferometer performance,
we consider two different quantities, namely
$\Delta \theta_{\text{ep}}$ and $\Delta \theta_{\text{QCR}}$.
We emphasize that the discussion in this section is specific
to the situation where the phase imprinting is based on the linear 
phase shifter Hamiltonian. Non-linear phase imprinting 
protocols modify how the various limits scale with the
number of (active) atoms~\cite{PRA2008_Hl:Choi,PRA2009_HL:Boixo,Nature2011_HL:Napolitano}.

The phase sensitivity
$\Delta \theta_{\text{ep}}$ is obtained through error propagation,
\begin{eqnarray}
\Delta \theta_{\text{ep}}=
\frac{\Delta N_{s}(t_1+t_2+t_3)}{|\partial_{\theta}N_s(t_1+t_2+t_3)|}
\label{defi_sensitivity_ep}.
\end{eqnarray}
Since it is evaluated at $t=t_1+t_2+t_3$,
it depends on all three steps of the interferometer sequence
as well as the initial state.
The fact that $\Delta \theta_{\text{ep}}$ is fully determined
by the characteristics of the side mode population 
makes it 
readily accessible to 
cold atom experiments.

A stringent limit on the parameter estimation is set by 
the 
phase sensitivity $\Delta \theta_{\text{QCR}}$, which
is derived from the quantum Cramer-Rao bound~\cite{RMP2018:Smerzi,PRL1994_QFI:Braunstein},
\begin{eqnarray}
\Delta \theta_{\text{QCR}}[|\Psi(t_1)\rangle,\hat{N}_s/2]=
\frac{1}{\sqrt{F_{Q}\left[|\Psi(t_1)\rangle,\hat{N}_s/2\right]}}.
\label{defi_sensitivity_QCR}
\end{eqnarray}
Here, $F_{Q}$ denotes the quantum Fischer information.
For the interferometer with linear phase shifter, the 
quantum Fischer information depends on 
$|\Psi(t_1) \rangle$ and the 
generator $\hat{N}_s/2$ that is associated with the
linear phase shifter Hamiltonian.
Importantly, the phase sensitivity $\Delta \theta_{\text{QCR}}$
is independent of the readout step. 
In general, one finds
\begin{eqnarray}
\Delta \theta_{\text{ep}} \ge \Delta \theta_{\text{QCR}}[|\Psi(t_1)\rangle,\hat{N}_s/2],
\label{defi_sensitivity_QCRBL1}
\end{eqnarray}
i.e., the 
quantum Cramer-Rao bound 
provides a lower bound for the
error propagation based sensitivity estimator.
For pure states $| \Psi(t_1) \rangle$
and linear phase imprinting generated
by $\hat{N}_s/2$, one finds~\cite{RMP2018:Smerzi}
\begin{eqnarray}
\Delta \theta_{\text{QCR}}
\left[|\Psi(t_1)\rangle,\hat{N}_s/2 \right]=\frac{1}{\Delta N_{s}(t_1)}.
\label{defi_sensitivity_QFI}
\end{eqnarray}
Ideally, one would like to operate in the regime where the quantity 
$\Delta \theta_{\text{ep}}/\Delta \theta_{\text{QCR}}$
is close to one, i.e., in the regime where the error propagation
based sensitivity $\Delta \theta_{\text{ep}}$ 
is as close as possible to the
best achievable phase sensitivity.

For comparison, we also report the Heisenberg limit
$\Delta \theta_{\text{HL}}$,
which we take to be defined in terms of the number
$N_s(t_1)$ of atoms
in the side modes at time $t_1$,
\begin{eqnarray}
\Delta \theta_{\text{HL}}=\frac{1}{N_{s}(t_1)};
\label{def_HL}
\end{eqnarray}
$N_s(t_1)$ can be thought of as the 
number of ``active atoms'' during the phase imprinting stage
of the SU(1,1) interferometer.
In this context, it is worthwhile mentioning that 
there exist a variety of definitions and interpretations of the Heisenberg
limit~\cite{PRA2008_Hl:Choi,PRL2010_HL:Anisimov,Nature2011_HL:Napolitano,OSA2018_HL:Tsarev,RMP2018:Smerzi,PRL1994_QFI:Braunstein,PRL2009_Hl:Pezz,PRA2009_HL:Boixo,PRL2016_SU11:Linnemann}.
Section~\ref{sec_results} shows that
the SU(1,1) interferometer with linear phase imprinting
allows for situations where the quantum Cramer-Rao bound based phase
sensitivity, which provides a strict lower bound, is
larger than the Heisenberg limit defined in Eq.~(\ref{def_HL}),
thereby
underpinning the notion that the Heisenberg limit, as defined in
Eq.~(\ref{def_HL}), should not be interpreted
as defining the ultimate or best achievable performance.

\subsection{Undepleted pump approximation}
\label{sec_upa}
The interferometer sequence 
introduced in Sec.~\ref{sec_IIB} has, in general,
to be modeled numerically.
Analytical results can, however, be obtained within the
undepleted pump approximation~(UPA)~\cite{PRL2015_SU11:Gabbrielli,PRL2016_SU11:Linnemann,PRL2017_Pump_up_SU11:Szigeti}, 
which replaces the operators $\hat{a}^{\dagger}_{0}$ 
and $\hat{a}_{0}$ by the square-root of
the mean number $\overline{N}_0$ of particles
in the $m=0$ mode
at time $t=0$,
\begin{eqnarray}
\overline{N}_{m}=\langle\Psi(0)|\hat{N}_{m}|\Psi(0)\rangle.
\end{eqnarray} 
The approximation is consistent with considering a
large reservoir of atoms 
in the $m=0$ pump mode.
Physically, the undepleted pump
approximation assumes that the majority of 
atoms occupies the $m=0$ pump mode.
This places restrictions
on the initial state and on the operating time $t_1+t_2+t_3$
of the interferometer, since only a small
fraction of the atoms should get pumped (i.e., scattered)
into the $m=+1$ and $m=-1$ side modes during the time evolution.

Dropping the constant energy shift
$-c(\overline{N}_0-1/2)/N-q$,
the spin Hamiltonian 
$\hat{H}_{\text{spin}}$ 
in the undepleted pump approximation reads
\begin{eqnarray}
\hat{H}_{\text{UPA}}(c,q)&=&\frac{2\overline{N}_{0}c}{N}\hat{K}_x \nonumber \\
&&+2 \left[
\frac{\overline{N}_0 c}{N}
\left( 1-\frac{1}{2 \overline{N}_0}\right)+q\right]\hat{K}_z,
\end{eqnarray}
where the operators 
$\hat{K}_{x}$, $\hat{K}_{y}$, and $\hat{K}_{z}$ are 
elements of the SU(1,1) group~\cite{PRA1986_SU11:Yurke},
\begin{eqnarray}
\hat{K}_{x}=
\frac{1}{2}
\left(
\hat{a}^{\dagger}_{+1}\hat{a}^{\dagger}_{-1}+
\hat{a}_{+1}\hat{a}_{-1}
\right),
\end{eqnarray}
\begin{eqnarray}
\hat{K}_{y}=\frac{1}{2 \imath}\left(\hat{a}^{\dagger}_{+1}\hat{a}^{\dagger}_{-1}-\hat{a}_{+1}\hat{a}_{-1}\right),
\end{eqnarray}
and
\begin{eqnarray}
\hat{K}_{z}=\frac{1}{2}\left(\hat{N}_{s}+1\right).
\end{eqnarray}
Since the Hamiltonian $\hat{H}_{\text{UPA}}$ can be written in terms of 
the elements of the SU(1,1) group, the resulting interferometer is
an SU(1,1) interferometer.
It is important to realize that $\hat{H}_{\text{UPA}}$
does not conserve the particle number.
In the context of photons, this is very natural.
In the context of spinor BECs as considered in this paper, this is not natural since 
the number of atoms is, neglecting one-, two-, and higher-body losses,
conserved.
We elaborate on this discussion in Sec.~\ref{sec_results_css}
and Appendix~\ref{appendix_css}. 

Looking ahead, we also define the simpler ``resonant'' Hamiltonian
$\hat{H}_{\text{UPA,r}}$, 
which assumes that the collisional and  Zeeman shifts cancel
each other, as a special case.
Setting 
$q$ in $\hat{H}_{\text{UPA}}$ to $q_c$,
\begin{eqnarray}
\label{eq_qc}
q_c=-\frac{\overline{N}_0 c}{N}
\left(1-\frac{1}{2 \overline{N}_0} \right),
\end{eqnarray}
we obtain
\begin{eqnarray}
\hat{H}_{\text{UPA,r}}(c)=\frac{2\overline{N}_{0}c}{N}\hat{K}_x.
\end{eqnarray}
Since $\overline{N}_0$ is assumed to be close to $N$ 
and $N$  is much greater than $1$,
we have 
$q_c \simeq -c$.

Our analytical results presented in Secs.~\ref{secIII}
and \ref{sec_results}
are
obtained for the standard three-step sequence of the SU(1,1)
interferometer, which is identical to the sequence
introduced in the previous section with $\hat{H}_{\text{spin}}$
replaced by $\hat{H}_{\text{UPA}}$.

\section{Solutions for SU(1,1) interferometer}
\label{secIII}
\subsection{Equations of motion}
\label{sec_eom}
To simulate the SU(1,1) interferometer sequence discussed in the previous
section, 
we work in the Heisenberg picture.
The equations of motion for the
time-dependent operators
$\hat{a}_{+1}$ and $\hat{a}_{-1}$
then read~\cite{PRA1986_SU11:Yurke}
\begin{eqnarray}
\label{eq_eom}
\imath \hbar\partial_{t}\hat{a}_{\pm1}(t)=
[\hat{a}_{\pm1}(t),\hat{H}_{\text{UPA}}(t)].
\end{eqnarray}
Solving the coupled linear equations implied
by Eq.~(\ref{eq_eom}), one obtains~\cite{PRA1986_SU11:Yurke}
\begin{eqnarray}
\begin{pmatrix} 
 \hat{a}_{+1}(t_1+t_2+t_3)  \\
 \hat{a}^{\dag}_{-1}(t_1+t_2+t_3)
\end{pmatrix} 
=
   \begin{pmatrix}
   \tilde{A} &
   \tilde{B} \\
   \tilde{B}^{*} &
  \tilde{A}^{*}
   \end{pmatrix} 
\begin{pmatrix} \hat{a}_{+1}(0)  \\ \hat{a}^{\dag}_{-1}(0)\end{pmatrix},
\end{eqnarray}
where the ``transfer matrix'' is constructed by applying 
three consecutive operations (one for each of the three
interferometer steps),
\begin{eqnarray}
\begin{pmatrix}
   \tilde{A}&
   \tilde{B}\\
   \tilde{B}^{*}&
  \tilde{A}^{*}
   \end{pmatrix}= \nonumber \\
\begin{pmatrix}
   A_{3} & B_{3} \\
   B^{*}_{3} &A^{*}_{3}
\end{pmatrix}\begin{pmatrix}
   e^{- \imath \theta /2}& 0 \\
  0 & e^{ \imath \theta /2}
   \end{pmatrix} \begin{pmatrix}
   A_{1} & B_{1} \\
   B^{*}_{1} &A^{*}_{1}
\end{pmatrix}. \label{interferometer_matrix}
\end{eqnarray}
Performing the matrix multiplication, one finds
\begin{eqnarray}
\tilde{A}=
A_{1}A_{3}e^{- \imath \theta/2}+B^{*}_{1}B_{3}e^{\imath \theta/2}
\end{eqnarray}
and
\begin{eqnarray}
\tilde{B}=
B_{1}A_{3}e^{-\imath \theta/2}+A^{*}_{1}B_{3}e^{ \imath \theta/2}.
\end{eqnarray}
Note that $\tilde{A}$ and $\tilde{B}$ depend on
$t_1$, $\theta$, and $t_3$ (recall
that $\theta$ depends on $t_2$);
for notational simplicity, these dependencies are not explicitly indicated. 
The quantities $A_{j}$ and $B_{j}$ depend on $t_{j}$. 
The next section
reports explicit expressions for 
$A_j$, $B_j$, $|\tilde{A}|^2$, and $|\tilde{B}|^2$ 
that are applicable to arbitrary
parameter combinations.

\subsection{General solution}
\label{sec_casei}
\begin{figure}[h]
  \centering
  \vspace*{1in}
\includegraphics[angle=0,width=0.45\textwidth]{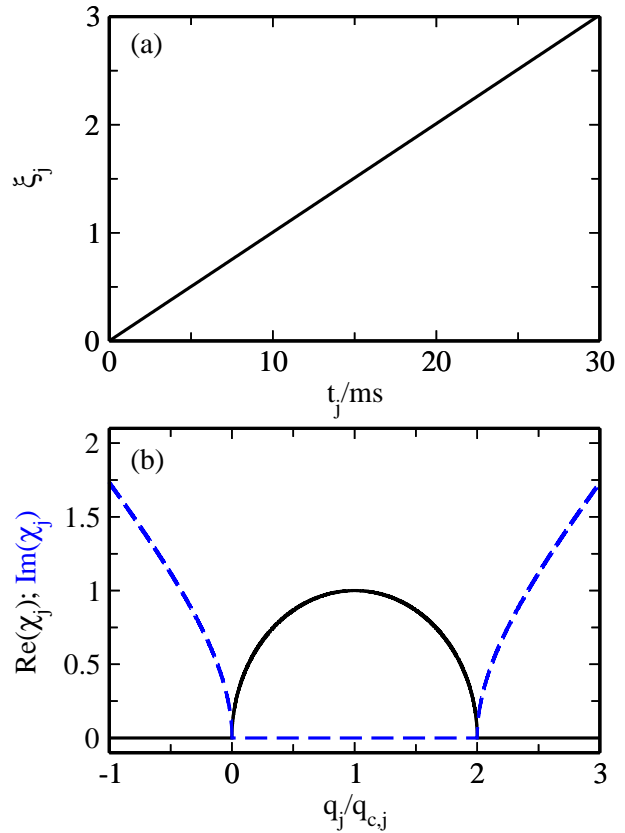}
\vspace*{0.in}
\caption{(Color online)
Dimensionless parameters
$\xi_j$ and $\chi_j$ that govern steps 1. and 3. of the 
interferometer sequence.
(a) The solid line shows $\xi_j$ as a function of $t_j$.
(b) The solid and dashed lines show the real and imaginary
parts of $\chi_j$ as a function of the
dimensionless parameter $q_j/q_{c,j}$.
The plots are made assuming a  
$^{23}$Na condensate with
$N=\overline{N}_0=10000$, 
$c_j/h=15.9956$~Hz, and $q_{c,j}/h = -15.9948$~Hz
(see Appendix~\ref{appendix_gp_equation} for details).}
\label{fig_param}
\end{figure}

Even though steps 1. and 3.
of the interferometer sequence
depend on six independent, experimentally
controllable parameters 
(namely 
$c_1$, $c_3$,
$q_1$, 
$q_3$, $t_1$, 
and $t_3$),
the solutions for
$A_j$ and $B_j$ ($j=1$ and 3) within the undepleted pump
approximation 
can be expressed in terms of 
four dimensionless parameters $\xi_1$, $\xi_3$,
$\chi_1$, and $\chi_3$, which are defined through
\begin{eqnarray}
\label{eq_xi}
\xi_j=\frac{\overline{N}_0 c_j t_j}{N \hbar}
\end{eqnarray}
and
\begin{eqnarray}
\label{eq_chi}
\chi_{j}=\sqrt{1-\left(1-\frac{q_j}{q_{c,j}}\right)^{2}\left(1-\frac{1}{2\overline{N}_{0}}\right)^{2}}.
\end{eqnarray}
Here, $q_{c,j}$ is given by Eq.~(\ref{eq_qc})
with $c$ replaced by $c_j$.
In what follows, we refer to $\xi_1$, $\xi_3$,
$\chi_1$, and $\chi_3$ as interferometer parameters.
As an example, Figs.~\ref{fig_param}(a) and \ref{fig_param}(b)
show the dependence of $\xi_j$ 
on the time $t_j$ and the dependence of 
the real and imaginary parts
of $\chi_j$ on the dimensionless parameter $q_j/q_{c,j}$ for a 
$^{23}$Na condensate
with $\overline{N}_0=N=10000$
(see Appendix~\ref{appendix_gp_equation} for details).

Using the parameters defined in
Eqs.~(\ref{eq_xi}) and (\ref{eq_chi}), $A_j$ and $B_j$ can be written as
\begin{eqnarray}
\label{eq_aj_general}
A_{j}=\cosh \left(\xi_{j}\chi_{j} \right)-\frac{\imath\sqrt{1-\chi_{j}^2}}{\chi_{j}}
\sinh \left( \xi_{j}\chi_{j} \right)
\end{eqnarray}
and
\begin{eqnarray}
\label{eq_bj_general}
B_{j}=-\frac{ \imath}{\chi_{j}}\sinh \left(\xi_{j}\chi_{j} \right).
\end{eqnarray}
Note that the interferometer performance may depend on
additional parameters that characterize
the initial state such as the initial
seeding fraction; $A_j$ and $B_j$ are, however, independent
of these additional parameters.
One finds
\begin{eqnarray}
|\tilde{A}|^2&=&\left(|A_{1}A_{3}|-|B_{1}B_{3}|\right)^2
+2|A_{1}A_{3}B_{1}B_{3}| \times \nonumber\\
&& \left[1+
\cos(\theta-\gamma_{A_{1}}-\gamma_{A_{3}}-\gamma_{B_{1}}+\gamma_{B_{3}})
\right]
\end{eqnarray}
and
\begin{eqnarray}
|\tilde{B}|^2&=&\left(|A_{1}B_{3}|-|A_{3}B_{1}|\right)^2
+2|A_{1}A_{3}B_{1}B_{3}| \times \nonumber\\
&& \left[1+
\cos(\theta-\gamma_{A_{1}}-\gamma_{A_{3}}-\gamma_{B_{1}}+\gamma_{B_{3}})
\right],
\end{eqnarray}
where the phases $\gamma_{A_{j}}$
and $\gamma_{B_j}$ are 
given by $\gamma_{A_{j}}=\arg(A_j)$ and $\gamma_{B_{j}}=\arg(B_j)$,
respectively.
It can be checked that 
both
$|A_j|^2 - |B_j|^2$ 
and
$|\tilde{A}_j|^2 - |\tilde{B}_j|^2$ are equal to $1$.
Appendix~\ref{appendix_property}
summarizes selected properties implied by 
Eqs.~(\ref{eq_xi})-(\ref{eq_bj_general}).

\subsection{Side mode population and associated
quantum fluctuation}
\label{sec_sidemodepopulation}
Armed with explicit expressions for $\hat{a}_{+1}(t)$,
$\hat{a}_{+1}^{\dagger}(t)$,
$\hat{a}_{-1}(t)$, 
and $\hat{a}_{-1}^{\dagger}(t)$ 
for all $t$ between $0$ and $t_1+t_2+t_3$,
the expectation value of the operator $\hat{N}_s$ and the corresponding
quantum fluctuation at any time $t$
can 
be calculated for any initial state $|\Psi(0) \rangle$, assuming the
undepleted pump approximation is valid.
We find
\begin{widetext}
\begin{eqnarray}
N_{s}(t_1+t_2+t_3)=
|\tilde{A}|^2 + |\tilde{B}|^2-1+
\left(|\tilde{A}|^2+|\tilde{B}|^2\right)\overline{N}_{s}+2\left(\tilde{A}\tilde{B}^{*}\overline{P}+\mbox{c.c.}\right)\label{general_Ns}
\end{eqnarray}
and
\begin{eqnarray}
&&\Delta N_{s}(t_1+t_2+t_3)
=2|\tilde{A}\tilde{B}|\times\nonumber\\
&&\sqrt{1+\overline{N}_{s}+2\overline{ \mbox{Cov} ( \hat{P}^{\dag}, \hat{P}) }
+\left[e^{2\imath \left(\gamma_{\tilde{A}}-\gamma_{\tilde{B}}\right)}\left(\overline{\Delta P}\right)^{2}+\mbox{c.c.}\right]+2 \left[e^{\imath \left(\gamma_{\tilde{A}}-\gamma_{\tilde{B}}\right)}\left(\overline{P}+\overline{ \mbox{Cov} ( \hat{N}_{s}, \hat{P}) }\right)+\mbox{c.c.}\right]{\cal{I}}+\left(\overline{\Delta N_{s}}\right)^{2}{\cal{I}}^{2}},~
\label{general_delta_Ns}
\end{eqnarray}
\end{widetext}
where $\gamma_{\tilde{A}}=\arg(\tilde{A})$ and $\gamma_{\tilde{B}}=\arg(\tilde{B})$.
The quantity ${\cal{I}}$ is independent of the properties of
the initial state,
\begin{eqnarray}
{\cal{I}} = 
\frac{|\tilde{A}|^2+|\tilde{B}|^2}{2|\tilde{A} \tilde{B}|}.
\end{eqnarray}
In Eqs.~(\ref{general_Ns}) and (\ref{general_delta_Ns}),
$\overline{N}_s$ is the number of atoms 
in the side modes at time zero
and $\overline{P}$,
\begin{eqnarray}
\label{def_bar_P}
\overline{P}=\langle\Psi(0)|\hat{P}|\Psi(0)\rangle,
\end{eqnarray} 
is defined in terms of the 
``pair annihilation operator'',
\begin{eqnarray}
\hat{P}=\hat{a}_{+1}\hat{a}_{-1}.
\end{eqnarray} 
The quantities
$\overline{\Delta N}_{s}$
and $\overline{\Delta {P}}$
denote the quantum
fluctuations associated with $\overline{N}_s$
and $\overline{P}$, respectively,
\begin{eqnarray}
\overline{\Delta {O}} = \Delta O(0)
\end{eqnarray}
(here, $O$ denotes an arbitrary observable).
Last, the quantity $\overline{\mbox{Cov}(\hat{O}_1, \hat{O}_2)}$ denotes the covariance 
of the operators $\hat{O}_1$ and $\hat{O}_2$ at time
zero,
\begin{eqnarray}
\overline{ \mbox{Cov} ( \hat{O}_1, \hat{O}_2) }=
\langle \Psi(0) | \hat{O}_1 \hat{O}_2 | \Psi(0) \rangle
- \nonumber \\
\langle \Psi(0) | \hat{O}_1  | \Psi(0) \rangle
\langle \Psi(0) | \hat{O}_2 | \Psi(0) \rangle.
\end{eqnarray}

As written,
Eqs.~(\ref{general_Ns}) and 
(\ref{general_delta_Ns}) apply
to an arbitrary pure initial state $|\Psi(0) \rangle$.
These equations also apply to an 
initial mixed state
$\hat{\rho}(0)$,
provided
$\overline{O}$, $\overline{\Delta O}$,
and
$\overline{ \mbox{Cov} (\hat{O}_1, \hat{O}_2) }$ are generalized.
For example,
$\overline{O}$ would be defined as $\mbox{Tr}[\hat{\rho}(0) \hat{O}]$
and analogous generalizations
would apply for the other expectation values.

As already discussed earlier,
$|\partial_{\theta} N_s(t_1+t_2+t_3)|$ and $\Delta N_s(t_1+t_2+t_3)$ govern the 
phase sensitivity $\Delta \theta_{\text{ep}}$.
It follows from
Eqs.~(\ref{general_Ns}) and (\ref{general_delta_Ns}) that the interferometer
performance depends on two 
aspects:
(i) the initial state through the
quantities
$\overline{N}_s$, $\overline{\Delta {N}}_{s}$, 
$\overline{P}$, 
$\overline{\Delta {P}}$,
$\overline{ \mbox{Cov} ( \hat{N}_s, \hat{P}) }$, and
$\overline{ \mbox{Cov} ( \hat{P}^{\dagger}, \hat{P}) }$; and
(ii) the interferometer device through $\tilde{A}$ and $\tilde{B}$.
Recall, $\tilde{A}$ and $\tilde{B}$ are, within the undepleted pump
approximation, fully determined by the 
five dimensionless parameters $\xi_{1}$, $\xi_{3}$, $\chi_{1}$, $\chi_{3}$, and
$\theta$.
Importantly, the first term on the right hand side of 
Eq.~(\ref{general_Ns}) and 
the first term inside the square root sign of Eq.~(\ref{general_delta_Ns})
depend only on the interferometer device while all other terms
``mix'' the interferometer device and the initial state.

We can also look at the quantum Cramer-Rao bound $1/\Delta N_s(t_1)$.
Equations~(\ref{general_Ns}) and (\ref{general_delta_Ns}) 
yield
$N_s(t_1)$ and $\Delta N_s(t_1)$
if $\tilde{A}$, $\tilde{B}$, $\gamma_{\tilde{A}}$, and $\gamma_{\tilde{B}}$ 
are replaced
by $A_1$, $B_1$, $\gamma_{A_{1}}$, and $\gamma_{B_{1}}$, respectively.
It follows that
$N_s(t_1)$
depends,
within the undepleted pump approximation, on the initial state
only through $\overline{N}_s$ and $\overline{P}$, i.e., the initial seeding and
the initial ``pair correlation''.
If $\overline{N}_s$ and $\overline{P}$ are zero,
$N_s(t_1)$ grows exponentially with increasing $|\xi_1 \chi_1|$
if $\chi_1$ is real. Maximal growth is obtained for $q_1=q_{c,1}$
(corresponding to $\chi_1=1$), with a growth
rate of $\overline{N}_0 c_1 / (N \hbar)$.
The regime where $N_s(t_1)$ grows exponentially is 
referred to as dynamical instability~\cite{NP2012_Amplification:Hamley}.
Since the fluctuation $\Delta N_s(t_1)$ 
depends
on the initial state, the quantum Cramer-Rao bound as well as
$\Delta \theta_{\text{ep}}$ can be
controlled, at least partially, by
adjusting the initial state.

\subsection{Special cases}
\label{sec_caseii}
The solutions presented in Secs.~\ref{sec_casei}
and \ref{sec_sidemodepopulation}
simplify significantly for the resonant
case, i.e., when $q_j$ is set to $q_{c,j}$
and $\hat{H}_{\text{UPA}}$ reduces
to $\hat{H}_{\text{UPA,r}}$.
Columns 1 and 2
of Table~\ref{tab_eom_results} summarize selected expressions for two resonant
cases
($\chi_1=\chi_3=1$), namely the
resonant symmetric interferometer for which $\xi=\xi_1=\xi_3$ and the 
resonant asymmetric
interferometer for which $\xi_1 \ne \xi_3$.
In the former case, the interferometer is fully characterized by 
the dimensionless parameter $\xi$
and the phase $\theta$;
this case has been considered in 
Ref.~\cite{PRL2016_SU11:Linnemann} for the vacuum state.
In the latter case, the interferometer is fully characterized by the two
dimensionless parameters $\xi_1$ and $\xi_3$ and the phase $\theta$;
this case has been considered in Ref.~\cite{PRL2015_SU11:Gabbrielli}
for a class of density matrices.

The solutions also simplify notably for the non-resonant
symmetric interferometer for which
$\xi=\xi_1=\xi_3 \ne 1$ and $\chi=\chi_1=\chi_3$.
In this case, the interferometer is fully characterized by 
the two dimensionless parameters $\xi$ and $\chi$ as well as the
phase $\theta$.
This case has been considered in Ref.~\cite{PRA2018_SU11:P.D.Lett}
and selected expressions 
are summarized in column 3
of Table~\ref{tab_eom_results}.

\begin{widetext}

\begin{table}[h]
\caption{Summary of the solutions within the undepleted
pump approximation to the equations of motion for three special cases: 
resonant symmetric interferometer, 
resonant asymmetric interferometer, and 
non-resonant symmetric interferometer.}
\label{tab_eom_results}
\begin{tabular}{c|c|c}
  resonant symmetric & resonant asymmetric & non-resonant symmetric \\ 
$\xi_{1}=\xi_3=\xi$; $\chi_{1}=\chi_3=1$ & any $\xi_{j}$; $\chi_{1}=\chi_3=1$&
$\xi_{1}=\xi_3=\xi$; $\chi_{1}=\chi_3=\chi$ \\ \hline
$A_{j}=A=\cosh\xi $ &$A_{j}=\cosh\xi_{j} $&$A_{j}=A=\cosh \left(\xi\chi \right)-\frac{\imath\sqrt{1-\chi^2}}{\chi}
\sinh \left( \xi\chi \right)$      \\ 
$ B_{j}=B=-\imath \sinh\xi$&$ B_{j}=-\imath \sinh\xi_j$ &$ B_{j}=B=-\imath \sinh(\xi\chi)$    \\ 
$\gamma_{A_{j}}=\gamma_{A}=0$ &$\gamma_{A_{j}}=\gamma_A=0$ &$\gamma_{A_{j}}=\gamma_{A}$   \\
$\gamma_{B_{j}}=\gamma_{B}=-\mbox{sign}(\xi)\pi/2$ &      
 $\gamma_{B_{j}}=-\mbox{sign}(\xi_j)\pi/2$
 &$\gamma_{B_{j}}=\gamma_{B}$   \\
$ |\tilde{B}|^2=\sinh^2(2\xi)\cos^2(\theta/2)$
&$ |\tilde{B}|^2= \cosh^2 \xi_1 \sinh ^2 \xi_3 +\cosh^2 \xi_3 \sinh ^2 \xi_1 +$
&$|\tilde{B}|^2=2 |AB|^2 \left[1+\cos(\theta-2\gamma_{A})\right]$    \\ 
& $2 | \cosh \xi_1 \cosh \xi_3 \sinh \xi_1 \sinh \xi_3 | \cos( \theta - \gamma_{B_1} + \gamma_{B_3})$
& \\
\end{tabular}
\end{table}
\end{widetext}

The SU(1,1) interferometer has a ``time reversal symmetry''
when $N_s(t)$
and $\Delta N_s(t)$ return 
at $t=t_1+t_2+t_3$ to their initial 
values
$\overline{N}_s$ and $\overline{\Delta {N}}_s$.
For the symmetric interferometer (i.e., $\gamma_{A_{1}}=\gamma_{A_{3}}=\gamma_{A}$ and 
$\gamma_{B_{1}}=\gamma_{B_{3}}=\gamma_B$), 
$\tilde{B}$ goes to zero at $t=t_1+t_2+t_3$
for $\theta = \pi + 2\gamma_{A}$.
For this phase, we  have
(the time dependence of $\tilde{A}$ 
is indicated explicitly for clarity)
\begin{eqnarray}
N_{s}(t_1+t_2+t_3)|_{\theta=\pi+2\gamma_{A}}=
|\tilde{A}(t_1+t_2+t_3)|^2\overline{N}_{s}~~~~~\label{time_rev_Ns}
\end{eqnarray}
and
\begin{eqnarray}
\Delta N_{s}(t_1+t_2+t_3)|_{\theta=\pi+2\gamma_{A}}=
|\tilde{A}(t_1+t_2+t_3)|^2 \overline{\Delta N}_{s}.~~~~~ \label{time_rev_Ns}
\end{eqnarray}
Thus, the symmetric 
SU(1,1) interferometer with $\theta = \pi+2\gamma_{A}$
has a time reversal symmetry
if the
initial state has no seeding, i.e., if $\overline{N}_s$ and $\overline{\Delta {N}}_s$
are equal to $0$.
If the interferometer is not only symmetric but also resonant
(in this case, $\gamma_{A}=0$)
and if we consider $\theta=\pi$,
then $\tilde{A}(t_1+t_2+t_3)$ goes to 1.
Thus, the interferometer has time reversal symmetry 
even when the initial state has non-zero seeding.
An SU(1,1) interferometer that utilizes time reversal symmetry was realized 
experimentally in a spinor $^{87}$Rb 
BEC~\cite{PRL2016_SU11:Linnemann}.

\section{Explicit results for various initial states}
\label{sec_results}
This section considers two typical classes of initial
states $| \Psi(0) \rangle$:
pure Fock states  are discussed in Sec.~\ref{sec_results_pfs}
and coherent spin states  in Sec.~\ref{sec_results_css}.
Selected properties of these two initial states
are summarized in Table~\ref{parameters_initial_state}.
Even though our analytical results
within the undepleted pump approximation are derived
in the Heisenberg picture, this section takes the
view point that the initial state is propagated in time and that
the operators are time independent.

\begin{widetext}

\begin{table}[h]
\caption{Properties of the initial states
$| \Psi(0) \rangle$ considered in Sec.~\ref{sec_results}: 
vacuum state (VS), pure Fock state with single- and double-sided
seeding (PFS,S and PFS,D), and coherent spin state with single- and
double-sided seeding (CSS,S and CSS,D).
The results are obtained within the undepleted pump approximation.
}
\label{parameters_initial_state}
\begin{tabular}{c|ccccccc}
$|\Psi(0)\rangle$  & 
$\overline{N}_s$  &$\overline{\Delta {N}}_s$&$\overline{P}$&
$\overline{\Delta {P}}$ & $\overline{\mbox{Cov}(\hat{N}_s,\hat{P})}$ & 
$\overline{\mbox{Cov}(\hat{P}^{\dagger},\hat{P})}$\\ \hline
VS&  0 &0&0&0 &0&0      \\ 
PFS,S& $ \overline{N}_s $ &0 &0&0   &0&0   \\ 
PFS,D&  $\overline{N}_s $ &0 &0&0  &0& $\overline{N}_+ \overline{N}_-$  \\
CSS,S &  $ \overline{N}_s $&$ \sqrt{\overline{N}_s} $ &0 &0  &0&0 \\ 
CSS,D&  $ \overline{N}_s $ &$ \sqrt{ \overline{N}_s} $
&$(\overline{N}_{+1}\overline{N}_{-1})^{1/2} \exp( -\imath \overline{\theta})$&  0 &0&0  
\end{tabular}
\end{table}

\end{widetext}

\subsection{Pure Fock state}
\label{sec_results_pfs}
Let the initial state be a pure Fock state
(PFS) with 
$\overline{N}_m$ atoms in mode $m$,
\begin{eqnarray}
|\Psi(0)\rangle=|\overline{N}_{+1},\overline{N}_{0},\overline{N}_{-1}\rangle\nonumber.
\end{eqnarray}
This initial state describes a system with $\overline{N}$ atoms,
where $\overline{N}=\overline{N}_{-1}+\overline{N}_0+\overline{N}_{+1}$.
We refer to 
$|0,\overline{N}_0,0\rangle$
as  ``vacuum  state'' (VS)~\cite{PRL2016_SU11:Linnemann}. 
The naming originates from the fact that the
side modes, sometimes also referred to as probe modes, are initially empty. 
The vacuum state can be interpreted as a special case
of a pure Fock state
or a special case of a coherent spin state (see Sec.~\ref{sec_results_css}). 
The spin mixing dynamics during step 1. of the
interferometer sequence can evolve the vacuum state to 
a state with significant entanglement~\cite{PRL2016_SU11:Linnemann}.
If one or both of the side modes contain 
non-zero occupation at time $t=0$,
we refer to the initial state as seeded Fock state.
Single-sided seeding is realized if $\overline{N}_{+1}$ 
or $\overline{N}_{-1}$ is non-zero 
and
double-sided seeding if $\overline{N}_{+1}$ 
and $\overline{N}_{-1}$ are non-zero.
We refer to the resulting states as 
pure Fock state with single-sided seeding (``PFS,S'') 
and double-sided seeding (``PFS,D''), respectively.

To calculate the phase sensitivity
$\Delta \theta_{\text{ep}}$, we need to determine
the expectation value of $\hat{N}_s$ and its quantum
fluctuation at time $t_1+t_2+ t_3$. Using 
the results for the pure
Fock state with double-sided seeding 
from Table~\ref{parameters_initial_state},
we find
\begin{eqnarray}
N_s(t_1+t_2+t_3)=\nonumber \\
|\tilde{A}|^2+|\tilde{B}|^2-1+\left(|\tilde{A}|^2+|\tilde{B}|^2\right)\overline{N}_{s}
\label{PFS_NS}
\end{eqnarray}
and
\begin{eqnarray}
\Delta N_{s}(t_1+t_2+t_3)=2|\tilde{A}\tilde{B}|\sqrt{1+\overline{N}_s+2\overline{N}_{+1}\overline{N}_{-1}}.~~~
\label{PFS_delta_NS}
\end{eqnarray}
Equations~(\ref{PFS_NS}) and (\ref{PFS_delta_NS}) show
that initial seeding leads to an
enhancement of the number of atoms
in the side modes at the end of the interferometer
sequence (i.e., speeds up the dynamics) and 
also enhances the quantum fluctuations. 
Physically, this can be interpreted as being due
to Bose enhancement as a consequence
of the non-zero initial seeding $\overline{N}_s$.
The quantum fluctuation
$\Delta N_s(t_1+t_2+t_3)$ does not only depend on $\overline{N}_s$ but,
for double-sided seeding, also on
the actual distribution of the atoms among the two side modes, i.e.,
the value of $\overline{N}_{+1} \overline{N}_{-1}$.
As shown in Eq.~(\ref{PFS_fs}), a non-zero $\overline{N}_+ \overline{N}_-$ leads
to a degradation of the phase sensitivity of the SU(1,1) interferometer.

\begin{figure}[t]
  \centering
  \vspace*{+1in}
\includegraphics[angle=0,width=0.4\textwidth]{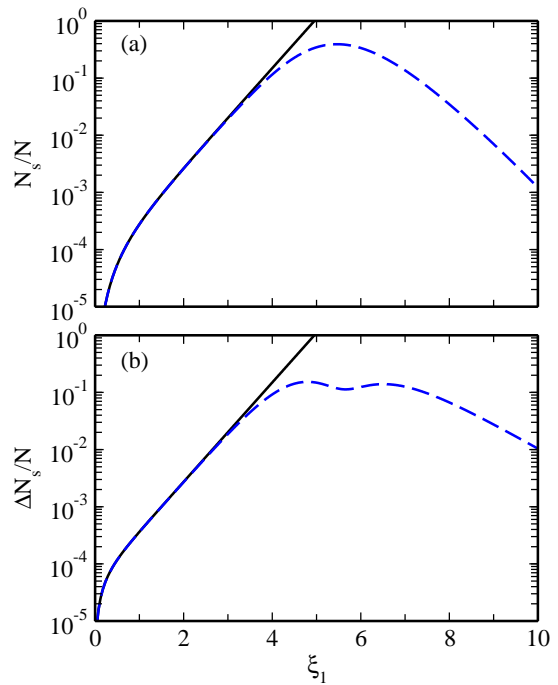}
\vspace*{0.0in}
\caption{(Color online)
Benchmarking the undepleted pump approximation
for the initial vacuum state $|0,\overline{N}_0,0\rangle$.
The solid lines show the undepleted pump approximation results 
for (a) $N_s$ [Eq.~(\ref{PFS_NS})]
and (b) $\Delta N_s$ [Eq.~(\ref{PFS_delta_NS})]
as a function of $\xi_1$ for $\chi_1=1$.
For comparison,
the dashed lines show our exact numerical results for the 
full spin Hamiltonian $\hat{H}_{\text{spin}}(c_1,q_1)$
with $N=\overline{N}_0=10000$, $c_1/h=15.9956$~Hz and $q_1/h=-15.9948$~Hz
(these are the same parameters as used 
in Fig.~\ref{fig_param}).
The agreement is good for $\xi_1 \lesssim 4$.
   }\label{fig_dynamics}
\end{figure} 

As an example,
Fig.~\ref{fig_dynamics} compares
our analytical results obtained within the undepleted pump
approximation (solid lines) with the results obtained by evolving the
initial state under
the full spin Hamiltonian $\hat{H}_{\text{spin}}$,
Eq.~(\ref{eq_hspin}), through exact
diagonalization (dashed lines) 
for a $^{23}$Na condensate with $\overline{N}_0=10000$.
It can be seen that the undepleted pump approximation
captures the time dependence of
$N_s$ [see Fig.~\ref{fig_dynamics}(a)]
and
$\Delta N_s$ [see Fig.~\ref{fig_dynamics}(b)]
well for $\xi_1 \lesssim 4$ 
(for the parameters employed, this corresponds to $t_1 \lesssim 40$~ms).
For $\xi_1 \gtrsim 4$, the undepleted pump approximation results 
deviate not only quantitatively but, rather quickly, also qualitatively 
from the exact numerical results.
From a practical point of view, 40~ms are sufficient for
an interferometer experiment.
For an initial state with $\overline{N}_s \ne 0$,
the validity regime of the undepleted pump approximation
tends to be somewhat more restricted.

From Eq.~(\ref{PFS_NS}) 
one obtains
\begin{eqnarray}
\partial_{\theta}N_{s}(t_1+t_2+t_3)=-4|A_{1}A_{3}B_{1}B_{3}|(1+ \overline{N}_{s})&&\nonumber\\
\times\sin(\theta-\gamma_{A_{1}}-\gamma_{A_{3}}-\gamma_{B_{1}}+\gamma_{B_{3}}).&&\label{PFS_partial_NS}
\end{eqnarray}
Combining Eqs.~(\ref{PFS_delta_NS}) and (\ref{PFS_partial_NS}),
the phase sensitivity [see Eq.~(\ref{defi_sensitivity_ep})] 
for the pure Fock state takes the form
\begin{eqnarray}
\Delta \theta_{\text{ep,PFS}}=
\Delta \theta_{\text{ep,VS}}
f_{\text{PFS}}
\label{PFS_Sensitivity},
\end{eqnarray}
where $f_{\text{PFS}}$ and $\Delta \theta_{\text{ep,VS}}$
are defined through
\begin{eqnarray}
f_{\text{PFS}}=
\frac{\sqrt{1 + \overline{N}_s + 2\overline{N}_{+1}\overline{N}_{-1}}}
{1+\overline{N}_s}
\label{PFS_fs}
\end{eqnarray}
and
\begin{eqnarray}
\Delta \theta_{\text{ep,VS}}=~~~~~~~~~~~~~~~~~~~~~~~~~~~~~~~~~~~~~~~~~~~~~~~~~\nonumber\\
\frac{|\tilde{A}\tilde{B}|}{2|A_1A_3B_1B_3 
\sin(\theta-\gamma_{A_{1}}-\gamma_{A_{3}}-\gamma_{B_{1}}+\gamma_{B_{3}})|}, \label{VS_Sensitivity}
\end{eqnarray}
respectively.
The quantity $\Delta \theta_{\text{ep,VS}}$ 
depends on $\theta$ through the $\sin$ term
in the denominator and through the $\cos$ terms in
$| \tilde{A} \tilde{B}|$,
\begin{widetext}
\begin{eqnarray}
| \tilde{A} \tilde{B}| =
\sqrt{
s
+
t \left[  1 + \cos(\theta-\gamma_{A_{1}}-\gamma_{A_{3}}-\gamma_{B_{1}}+\gamma_{B_{3}})\right]
+
\frac{u}{2} \left[1+\cos(\theta-\gamma_{A_{1}}-\gamma_{A_{3}}-\gamma_{B_{1}}+\gamma_{B_{3}}) \right]^2 
}
,
\end{eqnarray}
\end{widetext}
where
\begin{eqnarray}
\label{eq_s}
s=
\left(|A_{1}A_{3}|-|B_{1}B_{3}|\right)^2
\left(|B_{1}A_{3}|-|A_{1}B_{3}|\right)^2
,
\end{eqnarray}
\begin{eqnarray}
\label{eq_t}
t= 2|A_{1}A_{3}B_{1}B_{3}| \nonumber \\
 \times\left[\left(|A_{1}A_{3}|-|B_{1}B_{3}|\right)^2+
\left(|B_{1}A_{3}|-|A_{1}B_{3}|\right)^2 \right]
,
\end{eqnarray}
and
\begin{eqnarray}
\label{eq_u}
u=
8|A_{1}A_{3}B_{1}B_{3}|^2
.
\end{eqnarray}
Since $f_{\text{PFS}}$ reduces to $1$ for the 
vacuum state $|0,\overline{N}_0,0\rangle$,
the phase sensitivity for the vacuum state is given by 
$\Delta \theta_{\text{ep,VS}}$.
In this case,
$\Delta \theta_{\text{ep,VS}}$ 
can be interpreted as a phase sensitivity; for other initial
pure Fock states, in contrast,
$\Delta \theta_{\text{ep,VS}}$ is not
by itself a phase sensitivity but a function
that, together with $f_{\text{PFS}}$, determines the phase sensitivity 
$\Delta \theta_{\text{ep,PFS}}$. 
Since $\Delta \theta_{\text{ep,VS}}$
is 
determined by the ``actual device'' or interferometer parameters
(i.e., it is
independent of the initial state)
and $f_{\text{PFS}}$ is determined by the initial state
(i.e., it is independent of the interferometer parameters),
the minimum of $\Delta \theta_{\text{ep,PFS}}$
is determined by independently minimizing 
$\Delta \theta _{\text{ep,VS}}$ and $f_{\text{PFS}}$.
In what follows, we first analyze $f_{\text{PFS}}$ 
and then $\Delta \theta_{\text{ep,VS}}$. 

Since $f_{\text{PFS}}$ reduces
to $1$ in the absence of initial seeding, i.e., for $\overline{N}_s=0$,
we refer to it as ``seeding factor''.
We find
\begin{eqnarray}
\sqrt{\frac{1}{\overline{N}_{s}+1}}
\leq f_{\text{PFS}} \leq
\sqrt{\frac{1}{2}
\left[1+\frac{1}{(\overline{N}_{s}+1)^2}\right]}
\leq1\label{PFS_fs_domain}.
\end{eqnarray}
For non-zero $\overline{N}_s$, $f_{\text{PFS}}$ is always smaller than one.
Equations~(\ref{PFS_fs}) and (\ref{PFS_fs_domain}),
which apply to arbitrary initial pure Fock states
(assuming the undepleted pump approximation is applicable),
show:
\begin{itemize}
\item Initial seeding 
decreases the absolute phase sensitivity and hence improves the
absolute interfero\-meter performance.
\item For a fixed finite $\overline{N}_{s}$ and fixed interferometer
parameters, 
initial single-sided seeding leads to the best
interferometer performance 
(smallest $f_{\text{PFS}}$) and initial balanced double-sided
seeding to the worst interferometer performance 
(largest $f_{\text{PFS}}$) due 
to the presence of the ``pair term'' $\overline{N}_{+1} \overline{N}_{-1}$ in 
Eq.~(\ref{PFS_fs}). Importantly though, 
even initial balanced double-sided
seeding improves the interferometer performance compared to that
for the vacuum  state.
\end{itemize}

\begin{widetext}

\begin{table}[h]
\centering
\caption{Explicit expressions for
$\min(\Delta\theta_{\text{ep,VS}})$
[Eq.~(\ref{eq_minthetanew})]
and the associated $(\theta_{\text{min}})_{\text{VS}}$
[Eq.~(\ref{eq_mindeltathetanew})]
for 
the resonant symmetric interferometer,
resonant asymmetric interferometer, and 
non-resonant symmetric 
interferometer.
The results are obtained within the undepleted pump approximation.}
\label{min_VS}
\begin{tabular}{c|ccc}
 & resonant symmetric & resonant asymmetric & non-resonant symmetric \\ \hline
$(\theta_{\min})_{\text{VS}}$&  $ \pi$&$ \arccos\left[-\tanh\left(2\min(\xi_{1},\xi_{3})\right)\coth\left(2\max(\xi_{1},\xi_{3})\right)\right] + \gamma_{B_1} - \gamma_{B_3}$ &$\pi+2\gamma_{A}$   \\
$\min(\Delta\theta_{\text{ep,VS}})$&  $\text{csch} (2\xi) $ &$\text{csch}
\left[ 2 \min (\xi_1,\xi_3) \right] $&$\left| \frac{|A|^2-|B|^2}{2AB} \right|$    \\ 
\end{tabular}
\end{table}

\end{widetext}  

\begin{figure}
  \centering
  \vspace*{0.in}
\includegraphics[angle=0,width=0.45\textwidth]{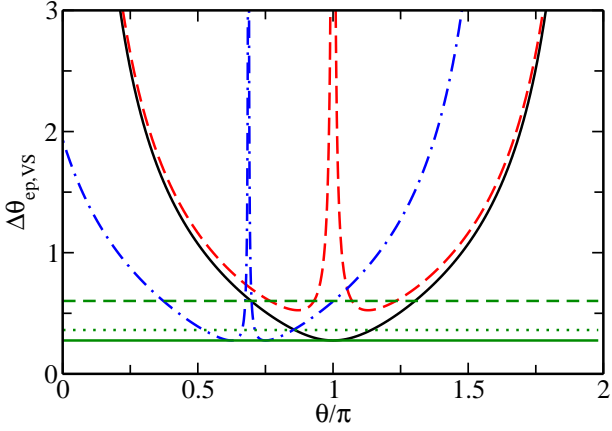}
\vspace*{0.in}
\caption{(Color online)
$\Delta \theta_{\text{ep,VS}}$
as a function of $\theta$ for three interferometers
that are characterized by the same $\xi_1$ and $\chi_1$
(namely, $\xi_1=\chi_1=1$) but different $\xi_3$ and $\chi_3$
for the case where the initial state is the vacuum state $|0,\overline{N}_0,0\rangle$.
The black solid curve
is for a resonant symmetric interferometer with
$\xi_3=\chi_3=1$,
the blue dash-dotted line is for a
non-resonant asymmetric 
interferometer
with
$\xi_3=3/2$ and $\chi_3=0$, and
the red dashed line is for a resonant asymmetric interferometer with
$\xi_3=7/10$ and $\chi_3=1$.
The minimum of the
phase sensitivity of the two former interferometers is the same
but the minimal value is reached at different angles.
The quantum Cramer-Rao bound $\Delta \theta_{\text{QCR}}$
(horizontal green solid line), 
the standard quantum limit
$[N_s(t_1)]^{-1/2}$ (horizontal green dashed line),
and
the Heisenberg limit $[N_s(t_1)]^{-1}$
(horizontal green dotted line)
are the same for all three interferometers.
}\label{fig_deltatheta_epvs}
\end{figure} 

We now analyze $\Delta \theta_{\text{ep,VS}}$. Due to the $\sin$ and $\cos$ dependence of 
$\Delta \theta_{\text{ep,VS}}$,
this quantity has a reflection symmetry around $\theta=\pi+\gamma_{A_{1}}+\gamma_{A_{3}}+\gamma_{B_{1}}-\gamma_{B_{3}}$.
Minimizing $\Delta\theta_{\text{ep,VS}}$ with respect to $\theta$
yields the best possible interferometer performance
for the SU(1,1) interferometer with pure Fock state input.
We find
\begin{eqnarray}
\label{eq_minthetanew}
\min(\Delta\theta_{\text{ep,VS}})=
\sqrt{
\frac{s+t+\sqrt{s(s+2t+2u)}}
{u}}
\end{eqnarray}
for
\begin{eqnarray}
\label{eq_mindeltathetanew}
(\theta_{\text{min}})_{\text{VS}}&=&
\mbox{arccos} 
\left(
- \frac{s+t+u-\sqrt{s(s+2t+2u)}}{t+u}
\right) \nonumber \\
&&+\gamma_{A_{1}}+\gamma_{A_{3}}+\gamma_{B_{1}}-\gamma_{B_{3}}.
\end{eqnarray}

Table~\ref{min_VS} summarizes explicit expressions for
$\min(\Delta\theta_{\text{ep,VS}})$ and 
$(\theta_{\min})_{\text{VS}}$ for the resonant symmetric,
resonant asymmetric, and non-resonant symmetric SU(1,1) interferometers. 
For symmetric interferometers,
the best performance (minimum of $\Delta \theta_{\text{ep,VS}}$)
is reached for $\theta=\pi+2 \gamma_{A}$,
i.e., the angle about which $\Delta \theta_{\text{ep,VS}}$
has a reflection symmetry.
The solid line in Fig.~\ref{fig_deltatheta_epvs} illustrates this for the
resonant symmetric interferometer
with $\xi_1=\chi_1=\xi_3=\chi_3=1$ ($\gamma_{A}=0$ and
$\gamma_{B_{1}}=\gamma_{B_{3}}=-\pi/2$),
implying that the best performance is reached for
$\theta=\pi$.
In this case, the minimum of the error
propagation based phase sensitivity coincides with the quantum
Cramer-Rao bound (horizontal green solid 
line in Fig.~\ref{fig_deltatheta_epvs}), which lies below the
standard quantum limit and below the Heisenberg limit
(horizontal green dashed and dotted
lines in Fig.~\ref{fig_deltatheta_epvs}).
This implies that the Heisenberg limit does, in this case, not
provide a stringent lower bound.
The quantum Cramer-Rao bound is also reached for
the non-resonant asymmetric interferometer with
$\xi_1=\chi_1=1$, $\xi_3=3/2$, and $\chi_3=0$
(blue dash-dotted line); in this case, however, the minimum
of the error propagation based
sensitivity is reached at a different angle, namely
at $\theta=0.624\pi$ and $0.75\pi$.
Returning to the 
asymmetric 
interferometer but 
considering the 
resonant 
case with $\xi_1=\chi_1=\chi_3=1$ and $\xi_3=7/10$
(red dashed line), we see that the
minimum of the error propagation
based phase sensitivity lies above 
that of the resonant symmetric interferometer with $\xi_3=1$.
In fact, for pure
Fock states 
the minimum of the error propagation based
phase sensitivity for the resonant interferometer
decreases monotonically with
increasing $\xi_3$ for $\xi_3 < \xi_1$ and then takes
a constant value for $\xi_{3}>\xi_1$
[see the black solid and blue dashed lines 
in Fig.~\ref{fig_min_sen}(a)].

With the properties of the quantities $f_{\text{PFS}}$ and
$\Delta \theta_{\text{ep,VS}}$ mapped out, we compare the
minimum of the phase sensitivity $\Delta \theta_{\text{ep,PFS}}$
and the corresponding quantum Cramer-Rao bound, i.e., we calculate the ratio
\begin{eqnarray}
\label{epoverQCRinPFS}
\frac{\min(\Delta\theta_{\text{ep,PFS}})}{\Delta\theta_{\text{QCR,PFS}}}=\frac{\min(\Delta\theta_{\text{ep,VS}})}{\Delta\theta_{\text{QCR,VS}}}\left(1+\frac{2\overline{N}_{+1}\overline{N}_{-1}}{1+\overline{N}_{s}}\right).~~~~
\end{eqnarray}
We do not have a general result for when this equation is equal to
one and when it is greater than one.
In the parameter range $0 < q_j/q_{\text{c},j}<2$, however,
a sufficient condition for 
the prefactor $\min(\Delta\theta_{\text{ep,VS}})/\Delta\theta_{\text{QCR,VS}}$
being equal to one is $\xi_3 \ge \xi_1$ and $\chi_3 \ge \chi_1$. Thus, in this parameter regime
(which includes a variety of non-resonant asymmetric
interferometers), a pure Fock state without seeding or
with single-sided seeding does reach the quantum Cramer-Rao bound
while a pure Fock state with double-sided seeding does not
reach the quantum Cramer-Rao bound.

\subsection{Coherent spin state}
\label{sec_results_css}
Let the initial state be a coherent spin state (CSS)~\cite{PR2012_Spinor_BEC:Ueda}.
Generally speaking,
coherent spin states with single- and double-sided seeding
are much easier to prepare experimentally than pure Fock
states with seeding.
Coherent spin states may be characterized as the
most classical of all quantum states~\cite{scully1999quantum}.
Thus, intuitively one might expect that coherent spin states
perform less well than pure Fock states for the same
interferometer parameters during steps 1. and 3. of the 
interferometer sequence. 
This section shows that coherent spin states
with double-sided seeding yield, in some cases,
a smaller error propagation based
phase sensitivity than pure Fock states.

Our analytical results within the undepleted pump approximation are derived for the
coherent spin state 
\begin{widetext}
\begin{eqnarray}
\label{eq_css_upa_state}
| \beta_{+1}, \beta_{-1} \rangle = \sum_{n_{+1}=0}^{\infty}
\sum_{n_{-1}=0}^{\infty}
\exp \left( - \frac{|\beta_{+1}|^2 + |\beta_{-1}|^2}{2} \right)
\frac{(\beta_{+1})^{n_{+1}} (\beta_{-1})^{n_{-1}}}{\sqrt{n_{+1}! n_{-1}!}} | n_{+1},n_{-1} \rangle,
\end{eqnarray}
\end{widetext}
where the complex numbers $\beta_{m}$
are written in terms of the 
initial atoms $\overline{N}_m$ in the side modes
and the initial phases $\overline{\theta}_m$
of the $m=+1$ and $m=-1$ modes, $\beta_m = (\overline{N}_m)^{1/2} \exp( \imath \overline{\theta}_m)$.
A derivation of this state is given in Appendix~\ref{appendix_css}.
Consistent with the fact that $\hat{H}_{\text{UPA}}$ does not conserve the number
of particles, this ``two-mode'' state ($|n_{+1},n_{-1}\rangle$ denotes a Fock state) is
characterized by a distribution of number of atoms.
The relative phase $\overline{\theta}$,
\begin{eqnarray}
\overline{\theta}=-(\overline{\theta}_{+1}+\overline{\theta}_{-1}),
\end{eqnarray}
of the state given in Eq.~(\ref{eq_css_upa_state})
is well-defined
in the case of double-sided seeding but not
in the case of
single-sided seeding or without seeding; the latter is
equal to the vacuum state.
In the case of initial double-sided seeding, 
 the interferometer
performance depends on the relative phase, thereby providing
another tuning knob.

\begin{widetext}
For the initial coherent spin state given in Eq.~(\ref{eq_css_upa_state}),
Eqs.~(\ref{general_Ns}) and (\ref{general_delta_Ns}) reduce to
\begin{eqnarray}
N_s(t_1+t_2+t_3)=|\tilde{A}|^2+|\tilde{B}|^2-1+
2 |\tilde{A} \tilde{B}| 
\left[ {\cal{I}} \,
\overline{N}_{s}+
2 \,
g(\overline{\theta},\gamma_{\tilde{A}},-\gamma_{\tilde{B}})
\sqrt{\overline{N}_{+1}\overline{N}_{-1}} \right]
\label{CSS_NS}
\end{eqnarray}
and
\begin{eqnarray}
\Delta N_{s}(t_1+t_2+t_3)=
2|\tilde{A} \tilde{B}| \sqrt{1+\left(1+{\cal{I}}^2\right)\overline{N}_s+4 \,{\cal{I}} \,
g(\overline{\theta},\gamma_{\tilde{A}},-\gamma_{\tilde{B}})
\sqrt{\overline{N}_{+1}\overline{N}_{-1}} 
},
\label{CSS_delta_NS}
\end{eqnarray}
\end{widetext}
respectively,
where the function $g(\overline{\theta},\gamma_{\tilde{A}},-\gamma_{\tilde{B}})$ is equal to zero
when $\overline{\theta}$ is not well defined 
(coherent spin state without seeding or with single-sided seeding)
and $g(\overline{\theta},\gamma_{\tilde{A}},-\gamma_{\tilde{B}})=\cos(\overline{\theta}-\gamma_{\tilde{A}}+\gamma_{\tilde{B}})$ when $\overline{\theta}$
is well defined 
(coherent spin state with double-sided seeding).
In the case of an initial state with double-sided seeding,
the function $g(\overline{\theta},\gamma_{\tilde{A}},-\gamma_{\tilde{B}})$
``mixes'' the properties of the initial state (through $\overline{\theta}$)
and the atual device (through $\gamma_{\tilde{A}}$ and
$\gamma_{\tilde{B}}$).

The next two sections separately discuss the interferometer performance
within the undepleted pump
approximation for coherent spin states with single- and double-sided seeding.
We have checked that our analytical 
results presented in the next two sections agree, up to terms of
order $1/N$,
with the numerical results for 
the full spin Hamiltonian $\hat{H}_{\text{spin}}$.
To make the comparisons, we used an initial coherent spin state
with fixed particle number [Eq.~(\ref{eq_appendix_css_1})]
in our exact diagonalization.

\subsubsection{Coherent spin state with single-sided seeding}
\label{sec_subsub1}
For an initial
coherent spin state with single-sided seeding, 
$N_s(t)$ is, for the same interferometer 
parameters, identical to that for a pure Fock state with single-sided seeding
[compare Eq.~(\ref{CSS_NS}) with Eq.~(\ref{PFS_NS})].
The quantum fluctuation $\Delta N_s(t)$ for the coherent spin state 
with single-sided seeding, in contrast, differs from that for
a pure Fock state with single-sided seeding since
${\cal{I}}$
is,
in general, 
non-zero
[compare Eq.~(\ref{CSS_delta_NS}) with Eq.~(\ref{PFS_delta_NS})].
Correspondingly,
the phase sensitivity for the coherent spin state
with single-sided seeding also differs from that for the
pure Fock state with single-sided seeding.
We find
\begin{eqnarray}
\label{CSS_S_Sensitivity}
\Delta \theta_{\text{ep,CSS,S}}=\Delta \theta_{\text{ep,VS}}
f_{\text{CSS,S}},
\end{eqnarray}
where 
\begin{eqnarray}
\label{CSS_S_Sensitivity_f}
f_{\text{CSS,S}}=
\frac{\sqrt{1+ \left(
1+
{\cal{I}}^2
\right) \overline{N}_s}}
{1+\overline{N}_s}.
\end{eqnarray}
For $\overline{N}_s=0$, Eq.~(\ref{CSS_S_Sensitivity}) reduces to 
$\Delta \theta_{\text{ep,VS}}$; this is in agreement with the fact that the
coherent spin state without seeding reduces to the vacuum  state.
The factor 
$f_{\text{CSS,S}}$ depends on the initial
state through $\overline{N}_s$ and the actual device
through ${\cal{I}}$.
The ${\cal{I}}^2$ term in 
round brackets under the square root in Eq.~(\ref{CSS_S_Sensitivity_f})
leads, for the same interferometer parameters, 
to a degradation of the best interferometer performance
for an initial coherent spin state with single-sided seeding
compared to that of a pure Fock state with single-sided seeding
[compare Eq.~(\ref{CSS_S_Sensitivity_f}) with Eq.~(\ref{PFS_fs})].
Importantly, the factor $f_{\text{CSS,S}}$ can take
values smaller than $1$. This implies that a coherent spin 
state with single-sided seeding can---for the same
interferometer parameters---perform better than an initial
vacuum state.

Since $\Delta \theta_{\text{ep,VS}}$ and 
$f_{\text{CSS,S}}$ both depend explicitly on $\theta$,
determining the best interferometer performance requires
that one minimizes the product 
$\Delta \theta_{\text{ep,VS}} f_{\text{CSS,S}}$, 
i.e., the two terms cannot be treated separately.
This differs from the pure Fock state case
considered in Sec.~\ref{sec_results_pfs}, where $\Delta \theta_{\text{ep,VS}}$
and $f_{\text{PFS}}$ could be minimized separately.
While the minimization of $\Delta \theta_{\text{ep,VS}} f_{\text{CSS,S}}$
can, in principle, be done analytically, the
resulting expression for the minimum of the
error propagation based phase sensitivity
is rather lengthy and not overly illuminating. The 
following examples illustrate selected characteristics
of the interferometer performance for
coherent spin states with single-sided seeding.

The minimum and maximum of 
${\cal{I}}$ are reached at $\theta=\gamma_{A_{1}}+\gamma_{A_{3}}+\gamma_{B_{1}}-\gamma_{B_{3}}$ and 
$\theta=\pi+\gamma_{A_{1}}+\gamma_{A_{3}}+\gamma_{B_{1}}-\gamma_{B_{3}}$, respectively,
\begin{eqnarray}
\min
{\cal{I}} 
|_{\theta=\gamma_{A_{1}}+\gamma_{A_{3}}+\gamma_{B_{1}}-\gamma_{B_{3}}}=\frac{t+2u}{\sqrt{2u(s+2t+2u)}}
\end{eqnarray}
and
\begin{eqnarray}
\max
{\cal{I}}
|_{\theta=\pi+\gamma_{A_{1}}+\gamma_{A_{3}}+\gamma_{B_{1}}-\gamma_{B_{3}}}=\frac{t}{\sqrt{2us}}.
\end{eqnarray}
Correspondingly, for fixed $\overline{N}_s$, $f_{\text{CSS,S}}$ takes
its minimum and maximum at these angles.
It is easy to check that
the $\theta$ dependence in $f_{\text{CSS,S}}$ and $\Delta \theta_{\text{ep,VS}}$
enters only through $\cos(\theta-\gamma_{A_{1}}-\gamma_{A_{3}}-\gamma_{B_{1}}+\gamma_{B_{3}})$.
Correspondingly, $f_{\text{CSS,S}}$ and $\Delta \theta_{\text{ep,VS}}$
have a reflection symmetry around $\theta=\pi+\gamma_{A_{1}}+\gamma_{A_{3}}+\gamma_{B_{1}}-\gamma_{B_{3}}$.
For this angle, the seeding factor $f_{\text{CSS,S}}$ 
diverges for symmetric interferometers and the phase sensitivity $\Delta \theta_{\text{ep,VS}}$ 
diverges for asymmetric interferometers. Thus,
$\Delta \theta_{\text{ep,CSS,S}}$ diverges at this angle for all interferometers.
This is consistent with the fact that
$\Delta N_s$ is finite and $\partial_{\theta} N_s$ is zero
for $\theta = \pi + \gamma_{A_{1}}+\gamma_{A_{3}}+\gamma_{B_{1}}-\gamma_{B_{3}}$.

\begin{figure}
  \centering
  \vspace*{0.in}
\includegraphics[angle=0,width=0.44\textwidth]{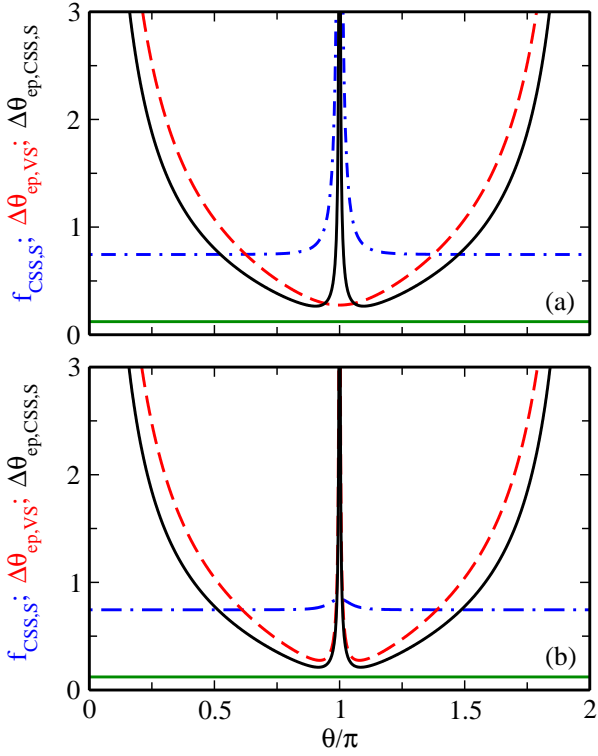}
\vspace*{0.in}
\caption{(Color online)
Analysis of the phase sensitivity 
for a coherent spin state
with single-sided seeding 
($\overline{N}_s=2$) as a function of $\theta$
for a resonant interferometer with $\chi_1=\chi_3=1$
($\gamma_{A_1}=\gamma_{A_3}=0$
and $\gamma_{B_1}= \gamma_{B_3}=-\pi/2$).
(a) Results for the resonant symmetric interferometer with 
$\xi_1=\xi_3=1$.
(b) Results for the 
resonant asymmetric 
interferometer
with $\xi_1=1$ and $\xi_3=3/2$.
The red dashed lines
show $\Delta \theta_{\text{ep,VS}}$
while the blue dash-dotted lines  show $f_{\text{CSS,S}}$.
The product of these two quantities yields the phase sensitivity
$\Delta \theta_{\text{ep,CSS,S}}$ (black solid lines). 
The horizontal green solid 
lines show the quantum Cramer-Rao bound.
   }\label{fig_css1}
\end{figure} 

As an example,
the black solid lines in Figs.~\ref{fig_css1}(a) and \ref{fig_css1}(b)
 show $\Delta \theta_{\text{ep,CSS,S}}$
for, respectively, a resonant symmetric  and a resonant asymmetric interferometer for an 
initial coherent spin state with single-sided seeding
($\overline{N}_s=2$).
Since $\gamma_{A_{1}}$ and $\gamma_{A_{3}}$ are equal to zero 
for resonant interferometers,
the reflection symmetry and divergence points 
of $\Delta \theta_{\text{ep,CSS,S}}$
are located at 
$\theta=\pi$.
For the resonant symmetric interferometer in Fig.~\ref{fig_css1}(a), the fact that ${\cal{I}}$ is minimized at the angle at
which $\Delta \theta_{\text{ep,VS}}$ is maximized
and that
${\cal{I}}$ diverges at the angle at
which $\Delta \theta_{\text{ep,VS}}$ is minimized
highlights that 
the quantities $\Delta \theta_{\text{ep,VS}}$ and 
$f_{\text{CSS,S}}$ ``compete'' when minimizing 
$\Delta \theta_{\text{ep,CSS,S}}$. 
As a consequence,
the smallest phase sensitivity is reached when neither
$f_{\text{CSS,S}}$ nor $\Delta \theta_{\text{ep,VS}}$ are minimized,
namely at $(\theta_{\text{min}})_{\text{CSS,S}}=0.904\pi$ and $1.096\pi$
for the example shown in Fig.~\ref{fig_css1}(a). 
For the resonant asymmetric interferometer in Fig.~\ref{fig_css1}(b),  
$\Delta \theta_{\text{ep,VS}}$ and $f_{\text{CSS,S}}$ take their maximum 
at $\theta=\pi$.
The minimum of $\Delta \theta_{\text{ep,CSS,S}}$ is lower than that
of $\Delta \theta_{\text{ep,VS}}$ but located, roughly, at the same angle.

The examples in Fig.~\ref{fig_css1} show that an initial coherent spin state
with single-sided seeding can improve the interferometer performance compared
to an initial vacuum state. 
While the minimum of the error propagation
based phase sensitivity for the coherent 
spin state with single-sided seeding is, for the examples shown in 
Fig.~\ref{fig_css1}, larger than the quantum Cramer-Rao bound
(horizontal green solid line), 
it is smaller than the quantum Cramer-Rao 
bound for the vacuum state (not
shown in Fig.~\ref{fig_css1}).
\begin{figure}
  \centering
  \vspace*{0.1in}
\includegraphics[angle=0,width=0.48\textwidth]{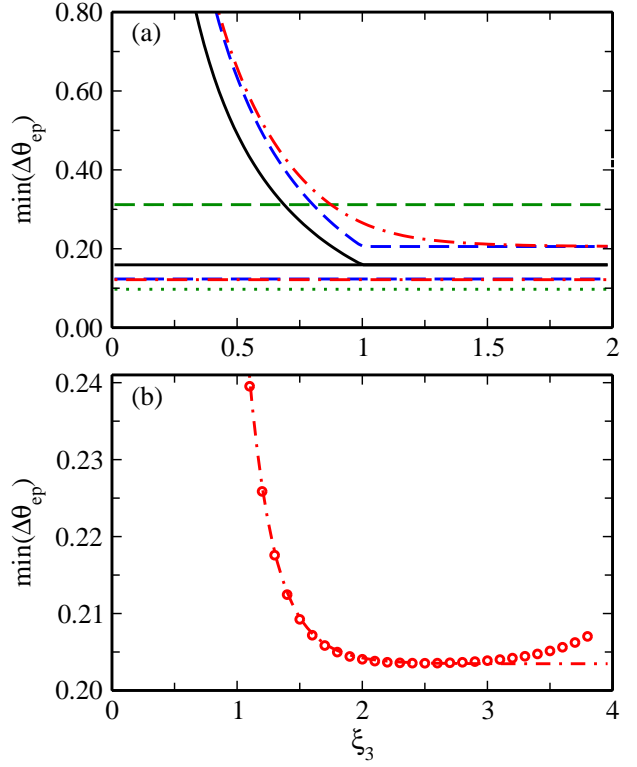}
\vspace*{0.in}
\caption{(Color online)
(a) Minimum of the phase sensitivity $\Delta\theta_{\text{ep}}$ for 
three different initial states 
with $\overline{N}_s=2$ 
for the resonant interferometer
with $\xi_1=\chi_1=\chi_3=1$
as a function of $\xi_3$.
Results are shown for
the pure Fock state with single-sided seeding (black solid line),
pure Fock state with double-sided seeding (blue dashed line), 
and 
coherent spin state with single-sided seeding (red dash-dotted line).
The three cases share the same standard
quantum limit (horizontal green dashed line) and 
the same Heisenberg limit 
$\Delta \theta_{\text{HL}}$ (horizontal green dotted line).
The quantum Cramer-Rao bound $\Delta\theta_{\text{QCR}}$
for the pure Fock state with single-sided seeding (horizontal black solid line),
the pure 
Fock state with double-sided seeding 
(horizontal
blue dashed line), and the coherent spin state
with single-sided seeding  (horizontal
red dash-dotted line) 
all lie above the Heisenberg limit
$\Delta\theta_{\text{HL}}$
(note that
the horizontal blue dashed and red dash-dotted lines nearly coincide).
(b) The symbols show the minimum 
$\mbox{min}(\Delta \theta_{\text{ep,CSS,S}})$ of the
phase sensitivity,
obtained by numerically determining the 
time evolution for the full spin Hamiltonian for a $^{23}$Na condensate
with 
$N=10000$, 
for the coherent spin state with single-sided seeding 
($\overline{N}_s=2$).
The red dash-dotted line from panel~(a) is reproduced
for comparison.
The deviations between the symbols and the dash-dotted line for 
$\xi_3 \gtrsim 3$
reflect the break-down of the undepleted pump approximation
in the long-time regime.
}\label{fig_min_sen}
\end{figure}

To highlight the dependence of the minimum of the phase sensitivity
$\Delta \theta_{\text{ep}}$ on the
initial state,
Fig.~\ref{fig_min_sen}(a) considers the resonant interferometer
with $\chi_1=\chi_3=\xi_1=1$ as a function
of $\xi_3$
for three different initial states:
pure Fock state with single-sided seeding (black solid line),
pure Fock state with double-sided seeding (blue dashed line),
and
coherent spin state with single-sided seeding (red dash-dotted line).
For $\xi_3 < \xi_1$, all three curves decrease monotonically with increasing
$\xi_3$, with 
$\Delta \theta_{\text{ep,CSS,S}} > \Delta \theta_{\text{ep,PFS,D}}>\Delta \theta_{\text{ep,PFS,S}}$
at each fixed $\xi_3$.
The phase sensitivities for the two pure Fock states are constant for 
$\xi_3>\xi_1$. For the coherent spin state, in contrast,
the minimum of the phase sensitivity for $\xi_3>\xi_1$
continues to decrease and approaches a constant in the
$\xi_3 \rightarrow \infty$ limit.
We emphasize that the decrease of $\Delta \theta_{\text{ep}}$ 
for $\xi_3 \gtrsim \xi_1$
is reproduced by our numerical calculations for the full
spin Hamiltonian for a $^{23}$Na BEC with $N=10000$
[symbols in Fig.~\ref{fig_min_sen}(b)].
However, for $\xi_3 \gg \xi_1$ the phase sensitivity
obtained for the full spin Hamiltonian deviates from the results
obtained within the undepleted pump approximation, underscoring
the fact that the long-time dynamics is not
described faithfully by the undepleted pump approximation.

Interestingly, the quantum Cramer-Rao bound for the three
states considered in Fig.~\ref{fig_min_sen}(a) all lie above the Heisenberg limit
(horizontal green dotted line).
This implies that the Heisenberg limit lies, in this case, below
the fundamental bound, i.e., the Heisenberg limit
can never be reached.

\subsubsection{Coherent spin state with double-sided seeding}
\label{sec_subsub2}
This section considers the interferometer performance 
for an initial coherent spin state with double-sided seeding.
Compared to the coherent spin state with single-sided seeding, the
relative phase $\overline{\theta}$ and the 
distribution of the atoms among the two
probe modes
(i.e., the product $\overline{N}_+ \overline{N}_-$)
provide additional
tuning knobs.

For an initial
coherent spin state with double-sided seeding,
the error propagation based phase sensitivity can be written as
\begin{eqnarray}
\label{sen_CSS_D}
\Delta\theta_{\text{ep,CSS,D}}=\Delta\theta_{\text{ep,VS}}f_{\text{CSS,D}},
\end{eqnarray}
where
\begin{widetext}
\begin{eqnarray}
\label{sen_CSS_D_f}
f_{\text{CSS,D}}=\frac{\sqrt{1+\left(1+
{\cal{I}}^2\right)\overline{N}_s+4 \,
{\cal{I}} \,
\cos (\overline{\theta}-\gamma_{\tilde{A}}+\gamma_{\tilde{B}})
\sqrt{\overline{N}_{+1}\overline{N}_{-1}}}}{\left|1+\overline{N}_{s}+
\left[2 \,
{\cal{I}} \,
\cos(\overline{\theta}-\gamma_{\tilde{A}}+\gamma_{\tilde{B}})+
2 \Delta \theta_{\text{ep,VS}}
\sin\left(\bar{\theta}-\gamma_{\tilde{A}}+\gamma_{\tilde{B}}\right)
\partial_{\theta}\left(\gamma_{\tilde{B}}-\gamma_{\tilde{A}}\right)
\right]
\sqrt{\overline{N}_{+1}\overline{N}_{-1}}\right|}.
\end{eqnarray}
\end{widetext}
It can be readily checked that $f_{\text{CSS,D}}$ reduces to 
$f_{\text{CSS,S}}$ if 
$\overline{N}_+ \overline{N}_-$ is equal to zero.
For non-zero $\overline{N}_+ \overline{N}_-$ and fixed interferometer
parameters, the terms in the numerator and denominator
of Eq.~(\ref{sen_CSS_D_f}) that contain
$\overline{N}_+ \overline{N}_-$ take,
depending on the value of 
$\overline{\theta} - \gamma_{\tilde{A}} + \gamma_{\tilde{B}}$,
positive or negative values:
The quantities ${\cal{I}}$ and $\Delta \theta_{\text{ep,VS}}$
are positive for all interferometer parameters;
$\partial_{\theta}(\gamma_{\tilde{B}} - \gamma_{\tilde{A}})$
and the $\sin$- and $\cos$-terms, in contrast, can take
positive or negative values.

\begin{figure}
  \centering
  \vspace*{0.1in}
\includegraphics[angle=0,width=0.47\textwidth]{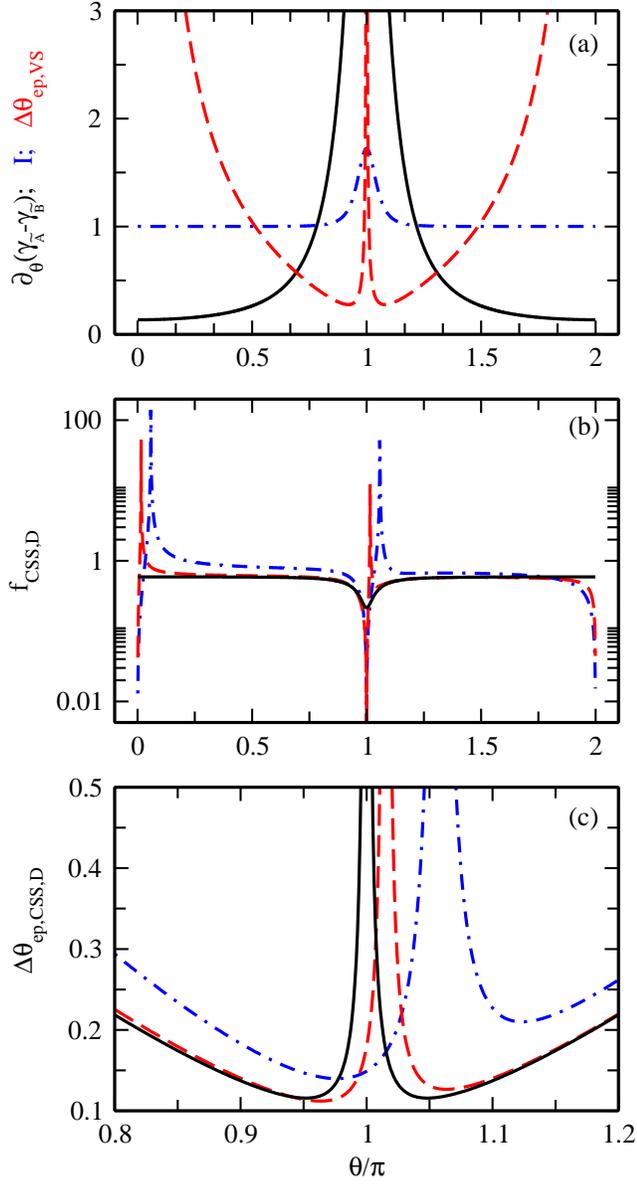}
\vspace*{0.in}
\caption{(Color online)
Analysis of error propagation based phase sensitivity 
for the resonant asymmetric interferometer with $\chi_1=\chi_3=\xi_1=1$ and
$\xi_3=3/2$ for an initial coherent spin state with 
$\overline{N}_s=2$. 
(a) The dashed, dash-dotted, and solid lines 
show 
$\Delta \theta_{\text{ep,VS}}$, ${\cal{I}}$, and 
$\partial_{\theta}(\gamma_{\tilde{B}} - \gamma_{\tilde{A}})$,
respectively,
as a function of $\theta$.
(b)
The solid, dashed, and dash-dotted lines show the quantity
$f_{\text{CSS,D}}$ 
for 
$\overline{\theta}=\pi/2$,
$\overline{\theta}=0.36 \pi$, and
$\overline{\theta}=0$,
respectively,
as a function of $\theta$ 
for a balanced initial
state with $\overline{N}_+=\overline{N}_-=1$.
Note the logarithmic scale of the vertical axis.
(c)
The solid, dashed, and dash-dotted lines show the 
phase sensitivity
$\Delta \theta_{\text{ep,CSS,D}}$
for 
$\overline{\theta}=\pi/2$,
$\overline{\theta}=0.36 \pi$, and
$\overline{\theta}=0$,
respectively,
as a function of $\theta$ for a balanced initial
state with $\overline{N}_+=\overline{N}_-=1$.}
\label{fig_css_new}
\end{figure}

To illustrate the interplay of the different terms that 
enter into $\Delta \theta_{\text{ep,CSS,D}}$,
we consider the resonant asymmetric interferometer
with $\chi_1=\chi_3=\xi_1=1$ and $\xi_3=3/2$.
The dashed, dash-dotted, and solid 
lines in Fig.~\ref{fig_css_new}(a)
show the quantities
$\Delta \theta_{\text{ep,VS}}$, ${\cal{I}}$, and 
$\partial_{\theta}(\gamma_{\tilde{B}} - \gamma_{\tilde{A}})$
as a function of $\theta$.
These quantities are fully determined by the
interferometer parameters, i.e., they 
are independent of the initial state.
As already discussed in the context of Fig.~\ref{fig_css1},
$\Delta \theta_{\text{ep,VS}}$ exhibits minima 
for $\theta$ just a bit larger and 
just a bit smaller than $\pi$.
For these angles,  
${\cal{I}}$ and 
$\partial_{\theta}(\gamma_{\tilde{B}} - \gamma_{\tilde{A}})$
take ``intermediate'' values (not maxima and not
minima).
Since 
$\Delta \theta_{\text{ep,CSS,D}}$ is
directly proportional to
$\Delta \theta_{\text{ep,VS}}$ and 
since 
$\Delta \theta_{\text{ep,VS}}$ 
also
enters through
the denominator of $f_{\text{CSS,D}}$,
$\Delta \theta_{\text{ep,CSS,D}}$ 
possesses a non-trivial dependence on $\theta$.

Choosing the balanced case with
$\overline{N}_+/\overline{N}_-=1$ as an example,
Fig.~\ref{fig_css_new}(b) shows
$f_{\text{CSS,D}}$ as a function of the phase shifter 
angle $\theta$ 
for various
initial phases $\overline{\theta}$ of the coherent spin state.
It can be seen that $f_{\text{CSS,D}}$ depends strongly 
on $\theta$ and $\overline{\theta}$:
$f_{\text{CSS,D}}$
changes by roughly four orders of magnitude 
for $\overline{\theta}=0.36 \pi$
and $\overline{\theta}=0$ and by less than an order of magnitude for
$\overline{\theta}=\pi/2$.
For the example shown, $f_{\text{CSS,D}}$ takes a
minimum at $\theta=\pi$ (for all $\overline{\theta}$ considered) and
a local minimum at $\theta=0$ and $2 \pi$ (for all $\overline{\theta}$ 
considered
except for $\overline{\theta}=\pi/2$).
These are exactly the angles at which $\Delta \theta_{\text{ep,VS}}$
diverges. 
Figure~\ref{fig_css_new}(c)
shows the error propagation
based phase sensitivity $\Delta \theta_{\text{ep,CSS,D}}$
for the same initial phases $\overline{\theta}$
as considered in Fig.~\ref{fig_css_new}(b).
It can be seen that the minimum of $\Delta \theta_{\text{ep,CSS,D}}$ is 
obtained for $\theta$ close to but not equal to $\pi$.

\begin{figure}
  \centering
  \vspace*{0.1in}
\includegraphics[angle=0,width=0.47\textwidth]{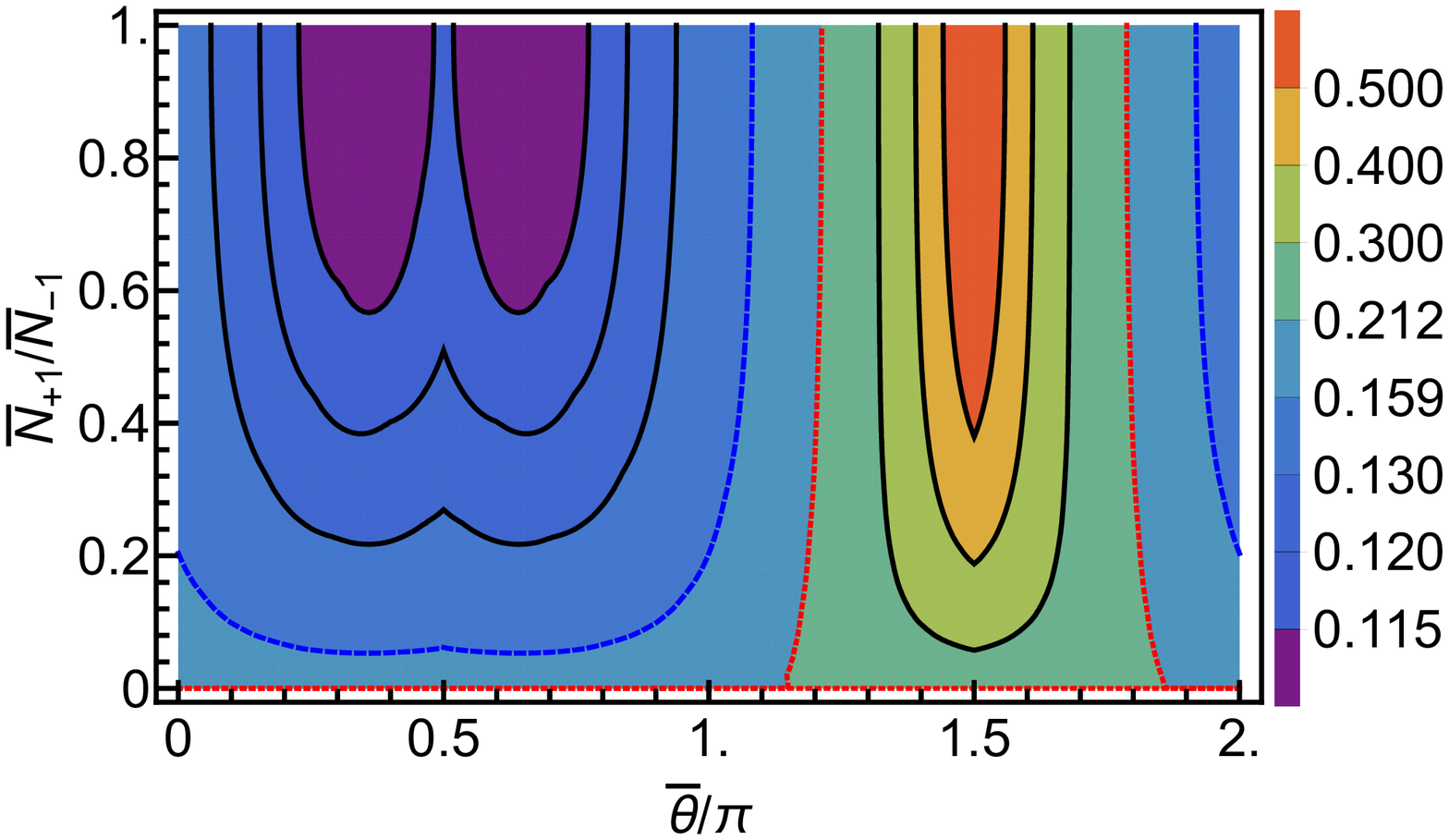}
\vspace*{0.in}
\caption{(Color online)
Minimum $\mbox{min}(\Delta \theta_{\text{ep,CSS,D}})$
of the  error propagation based phase sensitivity
for the resonant asymmetric interferometer with $\chi_1=\chi_3=\xi_1=1$ and
$\xi_3=3/2$ for an initial coherent spin state with 
$\overline{N}_s=2$
as functions of $\overline{N}_+/\overline{N}_-$ and $\overline{\theta}$. 
The legend on the right defines the color scheme of the contours.
The contours shown in blue and red
have the values 
$0.1592$ and $0.2120$, respectively.
The former is equal to 
$\min (\Delta \theta_{\text{ep,PFS,S}})$
and the latter to
$\min ( \Delta \theta_{\text{ep,PFS,D}})$.
}\label{fig_CSSD}
\end{figure}

Repeating the analysis for other $\overline{N}_+/\overline{N}_-$,
Fig.~\ref{fig_CSSD} shows the 
minimum $\mbox{min}(\Delta \theta_{\text{ep,CSS,D}})$
of the phase sensitivity 
as functions of $\overline{\theta}$ and $\overline{N}_+/\overline{N}_-$
for the same interferometer parameters as considered in 
Fig.~\ref{fig_css_new}.
For $\overline{N}_+/\overline{N}_-=0$, 
$\mbox{min}(\Delta \theta_{\text{ep,CSS,D}})$ is
independent of $\overline{\theta}$ and equal 
to 
$0.2120$;
this value agrees,
as it should, with the minimum of the error propagation based phase sensitivity
for the coherent spin state with single-sided seeding
(see the dash-dotted
line in Fig.~\ref{fig_min_sen} for $\xi_3=3/2$).
Figure~\ref{fig_CSSD} shows that the minimum
$\mbox{min}(\Delta \theta_{\text{ep,CSS,D}})$
of the error propagation based phase sensitivity 
can, depending on
the values of $\overline{N}_+/\overline{N}_-$ and $\overline{\theta}$,
be larger or smaller than 
$\mbox{min}(\Delta \theta_{\text{ep,CSS,S}})$.

It is  interesting to compare the performance
of the initial coherent spin state with double-sided
seeding with that for the initial pure Fock state
with single-sided seeding (and the same $\overline{N}_s$);
recall, among the pure Fock states,
the pure
Fock state with single-sided seeding
yields
the smallest
phase sensitivity $\Delta \theta_{\text{ep}}$
for fixed interferometer parameters.
For the interferometer parameters considered in
Fig.~\ref{fig_CSSD}, the
smallest phase sensitivity for the pure Fock state
is equal to $0.1592$.
Thus, the minimum of the phase sensitivity for the initial coherent spin state
with double-sided seeding
is, for a range of $\overline{\theta}$ and 
$\overline{N}_+/\overline{N}_-$, smaller than 
that for the pure Fock state with single-sided seeding.
This result is very encouraging as it points toward 
the possibility of achieving comparable or even better
phase or parameter estimates for initial coherent spin states
with double-sided seeding,
which are experimentally fairly straightforward to prepare,
than for initial pure Fock states with single-sided
seeding, which are experimentally rather challenging to prepare.

\begin{figure}
  \centering
  \vspace*{0.1in}
\includegraphics[angle=0,width=0.47\textwidth]{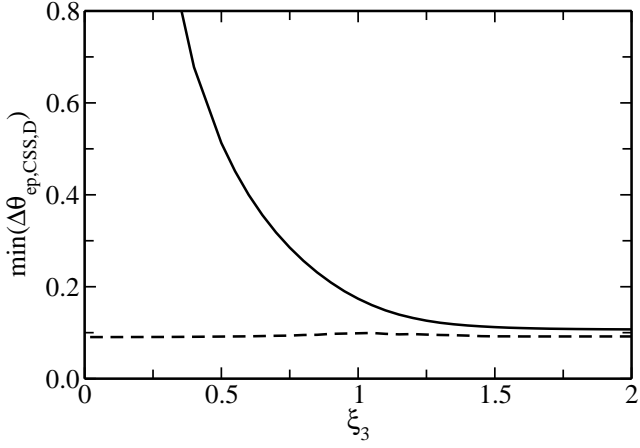}
\vspace*{0.in}
\caption{Interferometer performance for the resonant interferometer
with $\chi_1=\chi_3=\xi_1=1$
for a coherent spin state with 
double-sided seeding ($\overline{N}_s=2$).
The solid line
shows the minimum 
$\text{min} (\Delta \theta_{\text{ep,CSS,D}})$
of the error propagation
based phase sensitivity as a function of $\xi_3$;
note, the minimization is 
done by varying $\theta$, $\overline{\theta}$,
and
$\overline{N}_+/\overline{N}_-$.
For comparison, the dashed line shows the
quantum Cramer-Rao bound $\Delta \theta_{\text{QCR}}$ for the initial
states at which $\Delta \theta_{\text{ep,CSS,D}}$
takes its minimum.
}\label{fig_css_final}
\end{figure}

The solid line in Fig.~\ref{fig_css_final}
shows the minimum of the error propagation based
phase sensitivity
for an initial coherent spin state with $\overline{N}_s=2$ 
for the resonant 
interferometer with $\chi_1=\chi_3=\xi_1=1$ as a function
of $\xi_3$.
In this analysis, the minimum of the phase sensitivity is obtained
by minimizing $\Delta \theta_{\text{ep,CSS,D}}$ 
with respect to $\overline{\theta}$
and $\overline{N}_+ \overline{N}_-$
as well as the angle $\theta$.
For comparison, the dashed line shows the quantum Cramer-Rao bound,
calculated for each $\xi_3$ using the $\overline{\theta}$
and $\overline{N}_+ \overline{N}_-$
values 
that yield the smallest $\Delta \theta_{\text{ep,CSS,D}}$.
It can be seen that the error propagation based
phase sensitivity is closest to the quantum Cramer-Rao 
bound for the largest $\xi_3$ considered, i.e.,
for $\xi_3=2$.
By analyzing the dynamics for the
full spin Hamiltonian for a $^{23}$Na BEC with $N=10000$, we checked 
that the undepleted pump approximation
provides an accurate description for all $\xi_3$ values
considered in Fig.~\ref{fig_css_final}.
It is also instructive to compare
with the quantum Cramer-Rao bounds for the
double-sided and single-sided pure Fock states, which are
equal to 0.1233 and 0.1592, respectively, for $\xi_3 \ge 1$.
We find that the error propagation based
phase sensitivity for the coherent spin state
is lower than the quantum Cramer-Rao bound for the
pure Fock state for the same interferometer parameters
for $\xi_3 > 1.280$ and $\xi_3 > 1.054$, respectively.
This is interesting, since it indicates that 
the performance of the SU(1,1) interferometer, as 
quantified by the error propagation based phase
sensitivity, can ``beat'' the quantum Cramer-Rao bound
for the pure Fock state, assuming the same initial seeding 
$\overline{N}_s$ and the same interferometer parameters.

\section{Conclusions}
\label{sec_conclusion}
This work analyzed the performance
of a spin-1 Bose-Einstein condensate 
based interferometer
for
parameter combinations that
can be realized experimentally.
Within the undepleted pump approximation, which is
employed throughout this paper, 
the spinor BEC realizes an SU(1,1) interferometer,
in which the $m=0$ state of the 
$f=1$ hyperfine manifold
 serves as the pump and the $m=+1$ and $m=-1$ hyperfine 
states serve as the probe.
Although the interferometer itself, which consists of
the state preparation, phase imprinting, and read-out steps,
 has eight experimentally
tunable parameters ($t_1$, $q_1$, $c_1$,
$t_2$, $q_{\text{ps}}$, 
$t_3$, $q_3$, $c_3$),
it is characterized by five parameters
within the undepleted pump approximation:
two that describe the state preparation step,
one that describes the phase imprinting step, and
two that describe the read-out step.
The initial state adds additional degrees of freedom:
For pure Fock states the fraction of atoms in the three different
hyperfine states can be varied.
For coherent spin states with double-sided seeding, the initial
relative phase provides an additional tuning knob.
The validity regime of the undepleted
pump approximation limited our analysis to cases where the side mode
population of the initial state and the time propagated state are 
much smaller
than the population of the pump mode.

The dependence of the performance of the spinor-BEC based SU(1,1) interferometer on the seeding fraction and initial phase of coherent spin states was already investigated in Ref.~\cite{PRA2018_SU11:P.D.Lett} within the truncated Wigner approximation. As in our work, both phase insensitive amplifiers (single-sided seeding) and phase sensitive amplifiers (double-sided seeding) were considered.  Our analytical results, obtained within the undepleted pump approximation, confirm the results of Ref.~\cite{PRA2018_SU11:P.D.Lett} with regards to the role played by the initial state: (ai) Even a tiny seeding fraction has a non-negligible effect on the interferometer performance; correspondingly, an analysis of experimental interferometer results needs to account for possible imperfections of the initial state. 
(aii) The best interferometer performance 
$\mbox{min}(\Delta \theta_{\text{ep}})$
of a coherent spin state with single-sided seeding is obtained for a phase shifter angle that differs (typically just slightly) from that of the vacuum state; in fact, the angle at which the vacuum state performs best yields the worst performance for the coherent spin state with single-sided seeding. 
For the same interferometer parameters,
an initial coherent spin state with single-sided
seeding yields a lower $\mbox{min}(\Delta \theta_{\text{ep}})$ than an initial
vacuum state.
(aiii) The coherent spin state with double-sided seeding 
can perform better than the vacuum state 
and pure Fock states with single- and double-sided seeding
for appropriately chosen initial phases. This is a very encouraging result from the experimental point of view since coherent spin states with double-sided seeding can be realized fairly straightforwardly by first preparing a condensate in the 
$|f=1,m=0 \rangle$ state and by then applying a short radio-frequency pulse that transfers a fraction of the atoms into 
the $m= \pm 1$ states. The initial phases of the three hyperfine components can be controlled by introducing a finite variable detuning. Alternatively, microwave transitions that couple to the $f=2$ manifold can be used~\cite{PRA2006_MW_dressing:Gerbier,PRA2014_MW_dressing:Zhao,PRL2016_SU11:Linnemann}.

Additional findings of our work are: (bi) Within the undepleted pump approximation, analytical expressions that account for all three steps of the SU(1,1) interferometer sequence were presented for an arbitrary pure
initial state and analyzed for a subset of Fock and coherent spin states. We expect that these expressions, which can be straightforwardly implemented in Mathematica or other software packages, will aid the analysis of experimental results and serve as a benchmark for solutions that go beyond the undepleted pump approximation. (bii) Pure Fock states with single-sided seeding perform, assuming the same interferometer parameters, better than 
pure Fock states with double-sided seeding. 
Pure Fock states with double-sided seeding, in turn, perform 
better than the vacuum state. 
From a practical point of view, pure Fock states with non-zero seeding are rather fragile and hence challenging to work with experimentally. (biii) Parameter regimes where coherent spin states perform better than pure Fock states with single-sided seeding, in addition to performing better than the vacuum state [see point~(aiii) above], were identified (again, assuming the same interferometer parameters). This is encouraging since this finding underscores that the spin mixing dynamics can generate, starting from an initial coherent state that may be viewed as the most classical of all quantum states, useful “quantum-ness” or entanglement during the first stage of the interferometer sequence.

Last, we highlight a number of key findings that relate to the interferometer steps themselves.  (ci) 
For a coherent spin state with single- or double-sided seeding, 
the error propagation based sensitivity continues, for a wide range of parameters, to decrease for $\xi_3>\xi_1$, i.e., when $t_3>t_1$. This asymmetric behavior might be enhanced if one goes beyond the undepleted pump approximation. 
(cii) The quantum Cramer-Rao bound, which is fully 
determined by the state $| \Psi(t_1) \rangle$
that enters the linear phase imprinting step of the interferometer, provides the ultimate lower bound for the phase sensitivity; unfortunately, however, no general protocols for its direct experimental determination exist. The error propagation based phase sensitivity $\Delta \theta_{\text{ep}}$, in contrast,  depends on all three stages of the interferometer. Our calculations suggest that it is, in general, not possible to predict 
$\mbox{min}(\Delta \theta_{\text{ep}})$ by simply maximizing $N_s(t_1)$
or $\Delta N_s(t_1)$. 
While this is not unexpected, it highlights the interconnectedness of the various parameters. (ciii) The Heisenberg limit [taken to be given by $1/N_s(t_1)$] lies, for certain parameter combinations, below the quantum Cramer-Rao bound, indicating that one should, in general, work with the quantum Cramer-Rao bound and not with the Heisenberg limit. Since the observed behavior was verified by performing calculations for the full spin Hamiltonian, this conclusion is not an artifact of the undepleted pump approximation but valid more generally.

\section*{Acknowledgement}
We gratefully acknowledge discussions with Arne Schwettmann,
Qimin Zhang, and Shan Zhong.
Support by the National Science Foundation through
grant number
PHY-1806259
is gratefully acknowledged.
This work used
the OU
Supercomputing Center for Education and Research
(OSCER) at the University of Oklahoma (OU).

\appendix

\section{GP equation}
\label{appendix_gp_equation}

According to Eq.~(\ref{eq_cbar_to_c}),
the strength $c$ of the collision term
in the spin Hamiltonian is determined by $\overline{c}$ and
$\overline{n}$.
For $^{23}$Na, e.g., 
$a_0$ and $a_2$ are $48.91$~$a_{\text{bohr}}$ and 
$54.54$~$a_{\text{bohr}}$~\cite{PhysRevA.83.042704} ($a_{\text{bohr}}$ denotes the Bohr radius), 
respectively, leading
to $\overline{c}/h=1.54 \times 10^{-17}$~Hz$\,\text{m}^3$.
To determine the mean spatial density $\overline{n}$,
we treat a single-component $^{23}$Na BEC
within the mean-field Gross-Pitaevskii framework. 
To this end, we 
solve the Gross-Pitaevskii equation
\begin{eqnarray}
\label{eq_GP}
\Bigg[
- \frac{\hbar^2}{4 \mu} \nabla^2_{\vec{r}} + 
 \mu (\omega_x^2 x^2 + \omega_y^2 y^2 + \omega_z^2 z^2 )
+ \nonumber \\
\frac{2 \pi \hbar^2 (N-1) (a_0+2 a_2)}{3\mu}
|\psi_{\text{GP}}(\vec{r})|^2
\Bigg] \psi_{\text{GP}}(\vec{r}) = \nonumber \\
\epsilon \psi_{\text{GP}}(\vec{r}),
\end{eqnarray}
where 
$\epsilon$ denotes the chemical potential and 
$\psi_{\text{GP}}(\vec{r})$ the mean-field orbital,
which we take to be normalized to 1.
Given $\psi_{\text{GP}}(\vec{r})$,
the mean density $\overline{n}$ is given by
\begin{eqnarray}
\overline{n} = N \int |\psi_{\text{GP}}(\vec{r})|^4 d \vec{r}.
\end{eqnarray}

For an external harmonic trap with
angular frequencies 
$\omega_x=\omega_y = 2 \pi \times 166.277$~Hz and 
$\omega_z = 2 \pi \times 216.498$~Hz,
we find 
$\overline{n} = 1.04 \times 10^{18}$~m$^{-3}$,
$c_j/h=15.9956$~Hz
and $q_{c,j}/h = -15.9948$~Hz
for 
a
$^{23}$Na condensate with
$N=\overline{N}_0=10000$.
These are the values that are used to obtain the
results for $\hat{H}_{\text{spin}}$ shown in 
Figs.~\ref{fig_param}, \ref{fig_dynamics}, and \ref{fig_min_sen}(b).
We emphasize that the results obtained within the undepleted
pump approximation employ dimensionless parameters.
This implies that the undepleted
pump approximation results  shown
in Figs.~\ref{fig_dynamics}-\ref{fig_css_final} are 
applicable to a wide range of atomic species. The main
limitation is that the sign of $\xi_j$ is linked to the sign of
$c_j$, which can---in many cases---not be tuned.
For example, $c_j$ is positive for $^{23}$Na and negative for $^{87}$Rb.

\section{Properties of 
Eqs.~(\protect\ref{eq_xi})-(\protect\ref{eq_bj_general})}
\label{appendix_property}
Equations~(\ref{eq_xi})-(\ref{eq_bj_general}) imply the following:
\begin{enumerate}
\item The parameter $\xi_1$ and $\xi_3$
are real and their sign is determined by the
sign of the coupling strengths $c_1$ and $c_3$, respectively.
\item 
The parameters $\chi_{1}$ and $\chi_3$ are either purely real or purely imaginary.
\begin{itemize}
\item
For 
$-(2\overline{N}_{0}-1)^{-1}<q_{j}/q_{c,j} < (4\overline{N}_{0}-1)(2\overline{N}_{0}-1)^{-1}$ 
(this corresponds to $0 \lesssim q_j/q_{c,j} \lesssim 2$), 
$\chi_{j}$ is purely real. 
\item
For 
$q_{j}/q_{c,j} \leq -(2\overline{N}_{0}-1)^{-1}$ 
and $q_{j}/q_{c,j} \geq (4\overline{N}_{0}-1)(2\overline{N}_{0}-1)^{-1}$
(this corresponds to $q_j/q_{c,j} \lesssim 0$
and $q_j / q_{c,j} \gtrsim 2$), 
$\chi_{j}$ is purely imaginary,
with the imaginary part being greater than zero. In this case, it is
convenient to replace $\chi_{j}$ 
by $\imath |\chi_{j}|$.
Correspondingly,
it is convenient to replace
$\mbox{cosh}(\xi_j \chi_{j})$
in the expressions for $A_j$ and $B_j$ by $\mbox{cos}(\xi_j |\chi_{j}|)$
 and
to replace $\imath \mbox{sinh}(\xi_j \chi_{j}) /\chi_{j}$ by 
$\imath \mbox{sin}(\xi_j |\chi_{j}|)/|\chi_{j}|$. 
\end{itemize}
\item It follows that $A_1$ and $A_3$
are, in general, complex.
\item From points 1. and 2., it also follows that $B_{1}$ and $B_3$ are 
purely imaginary, implying that the phases $\gamma_{B_{j}}$ are 
equal to $\pi/2$ when the imaginary part of $B_j$ is positive and
$-\pi/2$ when the imaginary part is negative. 
\end{enumerate}

\section{Coherent spin state}
\label{appendix_css}
This appendix derives the expression for the
coherent 
spin state, Eq.~(\ref{eq_css_upa_state}),
used in Sec.~\ref{sec_results_css} within the undepleted pump
approximation.
The derivation starts with
the coherent three-mode spin state
$|\alpha_{+1},\alpha_{0},\alpha_{-1}\rangle$,
\begin{widetext}
\begin{eqnarray}
\label{eq_appendix_css_1}
|\alpha_{+1},\alpha_{0},\alpha_{-1}\rangle=
\underbrace{\sum_{n_{+1}=0}^N \sum_{n_{0}=0}^N \sum_{n_{-1}=0}^N}_
{n_{+1}+n_{0}+n_{-1}=N}
\sqrt{\frac{N!}{n_{+1}!n_{0}!n_{-1}!}}(\alpha_{+1})^{n_{+1}}(\alpha_{0})^{n_{0}}(\alpha_{-1})^{n_{-1}}|n_{+1},n_{0},n_{-1}\rangle,
\end{eqnarray} 
where the sums over the occupation numbers $n_{+1}$, $n_0$ and $n_{-1}$
are restricted such that the number $N$ of particles is fixed.
In Eq.~(\ref{eq_appendix_css_1}), we have
$\alpha_m=(\overline{N}_{m}/N)^{1/2} \exp(\imath\overline{\vartheta}_{m})$
and $\sum_{m=+1,0,-1}|\alpha_m|^2=1$.
This coherent three-mode spin state yields, when employed 
as an initial state for the time propogation under
the full spin Hamiltonian $\hat{H}_{\text{spin}}$,
results that agree up to order $1/N$ with the UPA results presented
in Secs.~\ref{sec_subsub1} and \ref{sec_subsub2}.
For example, this state is used to obtain the circles
in Fig.~\ref{fig_min_sen}(b).
When using Eq.~(\ref{eq_appendix_css_1}),
we define 
$\overline{\theta}=-(\overline{\vartheta}_{+1} + \overline{\vartheta}_{-1})$,
i.e., we set $\overline{\vartheta}_{0}$ to zero.
This does not pose any restrictions on our formulation since the results
are independent of the overall phase factor of the initial state.

The coherent three-mode spin state can alternatively be written
as
\begin{eqnarray}
\label{eq_appendix_css_2}
|\alpha_{+1},\alpha_{0},\alpha_{-1}\rangle=
\frac{1}{\sqrt{N!}} \left(
\alpha_{+1} \hat{a}_{+1}^{\dagger}+
\alpha_{0} \hat{a}_{0}^{\dagger}+
\alpha_{-1} \hat{a}_{-1}^{\dagger}
\right) ^N | \text{vac} \rangle,
\end{eqnarray} 
where $|\text{vac} \rangle$ denotes the vacuum state.
This is the ``true'' vacuum state that contains no
particles. It is distinct 
from the unseeded
Fock state $|0,N,0 \rangle$, which is referred to as vacuum state
throughout this paper in analogy with the photonic system.
Adding
$(N- \overline{N}_{+1}- \overline{N}_0- \overline{N}_{-1})
\exp(\imath \overline{\vartheta}_0 )$, which is equal to zero,
to the terms in the round brackets in Eq.~(\ref{eq_appendix_css_2}),
we find
\begin{eqnarray}
\label{eq_appendix_css_3}
|\alpha_{+1},\alpha_{0},\alpha_{-1}\rangle= 
\frac{N^{N/2} \exp( \imath N \overline{\vartheta}_0)}{\sqrt{N!}}
\times \nonumber \\
 \left[
1 + 
\frac{
\sqrt{\overline{N}_{+1}} \exp( \imath  \overline{\theta}_{+1}) 
 \hat{a}_{+1}^{\dagger}+
\sqrt{\overline{N}_{-1}} \exp( \imath \overline{\theta}_{-1})
 \hat{a}_{-1}^{\dagger}-
\overline{N}_{+1} - \overline{N}_{-1} +
\left( \sqrt{\overline{N}_{0}} \hat{a}_{0}^{\dagger}-\overline{N}_0 \right)
}{N}
\right] ^N | \text{vac} \rangle,
\end{eqnarray} 
where we defined $\overline{\theta}_{+1}= \overline{\vartheta}_{+1}
- \overline{\vartheta}_0$
and
$\overline{\theta}_{-1}= \overline{\vartheta}_{-1}
- \overline{\vartheta}_0$.
Considering the large $N$ limit and using the identity
\begin{eqnarray}
\lim_{N \rightarrow \infty} \left(
1+ \frac{A}{N} \right)^N = \exp (A),
\end{eqnarray}
we find
\begin{eqnarray}
\label{eq_appendix_css_4}
|\alpha_{+1},\alpha_{0},\alpha_{-1}\rangle 
\stackrel{N \rightarrow \infty}{\rightarrow}
\frac{N^{N/2}\exp( \imath N \overline{\vartheta}_0) 
}{\sqrt{N!}} 
\times \nonumber \\
\exp \left( -\overline{N}_{+1} - \overline{N}_{-1} \right)
\exp \left(
\sqrt{\overline{N}_{+1}} \exp( \imath  \overline{\theta}_{+1}) 
 \hat{a}_{+1}^{\dagger}+
\sqrt{\overline{N}_{-1}} \exp( \imath \overline{\theta}_{-1})
 \hat{a}_{-1}^{\dagger}
\right)
\exp \left(
\sqrt{\overline{N}_0} \hat{a}_0^{\dagger} - \overline{N}_0
\right)
| \text{vac} \rangle .
\end{eqnarray} 
Importantly, the right hand side of
Eq.~(\ref{eq_appendix_css_1}) 
is, in the large $N$ limit, identical to
Eq.~(\ref{eq_appendix_css_4}), i.e., Eq.~(\ref{eq_appendix_css_4})
is the coherent three-mode spin state for large $N$.

In the spirit
of the undepleted pump approximation, 
we now replace the operator $\hat{a}_0^{\dagger}$ in 
Eq.~(\ref{eq_appendix_css_4}) by $(\overline{N}_0)^{1/2}$.
This replacement has the following consequences:
(i) The term $\exp \left(
\sqrt{\overline{N}_0} \hat{a}_0^{\dagger} - \overline{N}_0
\right)$
goes to 1.
(ii) No atoms are created in the $m=0$ hyperfine state,
i.e., the $m=0$ mode of the three-mode state is
effectively being eliminated.
(iii) 
Expanding out the
exponential that contains the operators $\hat{a}_{+1}^{\dagger}$ 
and
$\hat{a}_{-1}^{\dagger}$, it can be seen that the state 
is a superposition of Fock states containing varying number of atoms;
this observation is closely related to point (ii) and also
implies that $N$ should now be interpreted as a parameter as
opposed to the actual atom number.
(iv) The state is no longer normalized to 1.
Restoring the normalization and using 
$\beta_m = (\overline{N}_m)^{1/2} \exp( \imath \overline{\theta}_m)$
(see Sec.~\ref{sec_results_css}), the right hand side of 
Eq.~(\ref{eq_appendix_css_4}) becomes
\begin{eqnarray}
\exp \left( \imath N \overline{\vartheta}_0 \right)
\exp \left(
-\frac{|\beta_{+1}|^2 + |\beta_{-1}|^2}{2}
\right)
\exp
\left(
\beta_{+1} \hat{a}_{+1}^{\dagger} + \beta_{-1} \hat{a}_{-1}^{\dagger}
\right) |\text{vac} \rangle.
\end{eqnarray}
\end{widetext}
Except for the overall phase $\exp( \imath N \overline{\vartheta}_0)$, 
which does not
have an effect on any of the observables, this is
the coherent spin state,
Eq.~(\ref{eq_css_upa_state}), used in our undepleted pump approximation
calculations in Sec.~\ref{sec_results_css}.
We emphasize that even though the $m=0$ mode has been effectively eliminated
from the formulation, this mode still serves as a
phase reference. This can be seen from the fact 
that  $\overline{\theta}_{+1}$
and $\overline{\theta}_{-1}$
are defined in terms of $\overline{\vartheta}_{+1}$
and
$\overline{\vartheta}_{-1}$, measured relative to $\overline{\vartheta}_0$.



\begin{thebibliography}{36}


\makeatletter
\providecommand \@ifxundefined [1]{%
 \@ifx{#1\undefined}
}%
\providecommand \@ifnum [1]{%
 \ifnum #1\expandafter \@firstoftwo
 \else \expandafter \@secondoftwo
 \fi
}%
\providecommand \@ifx [1]{%
 \ifx #1\expandafter \@firstoftwo
 \else \expandafter \@secondoftwo
 \fi
}%
\providecommand \natexlab [1]{#1}%
\providecommand \enquote  [1]{``#1''}%
\providecommand \bibnamefont  [1]{#1}%
\providecommand \bibfnamefont [1]{#1}%
\providecommand \citenamefont [1]{#1}%
\providecommand \href@noop [0]{\@secondoftwo}%
\providecommand \href [0]{\begingroup \@sanitize@url \@href}%
\providecommand \@href[1]{\@@startlink{#1}\@@href}%
\providecommand \@@href[1]{\endgroup#1\@@endlink}%
\providecommand \@sanitize@url [0]{\catcode `\\12\catcode `\$12\catcode
  `\&12\catcode `\#12\catcode `\^12\catcode `\_12\catcode `\%12\relax}%
\providecommand \@@startlink[1]{}%
\providecommand \@@endlink[0]{}%
\providecommand \url  [0]{\begingroup\@sanitize@url \@url }%
\providecommand \@url [1]{\endgroup\@href {#1}{\urlprefix }}%
\providecommand \urlprefix  [0]{URL }%
\providecommand \Eprint [0]{\href }%
\providecommand \doibase [0]{http://dx.doi.org/}%
\providecommand \selectlanguage [0]{\@gobble}%
\providecommand \bibinfo  [0]{\@secondoftwo}%
\providecommand \bibfield  [0]{\@secondoftwo}%
\providecommand \translation [1]{[#1]}%
\providecommand \BibitemOpen [0]{}%
\providecommand \bibitemStop [0]{}%
\providecommand \bibitemNoStop [0]{.\EOS\space}%
\providecommand \EOS [0]{\spacefactor3000\relax}%
\providecommand \BibitemShut  [1]{\csname bibitem#1\endcsname}%
\let\auto@bib@innerbib\@empty
\bibitem [{\citenamefont {Pezz\`e}\ \emph {et~al.}(2018)\citenamefont
  {Pezz\`e}, \citenamefont {Smerzi}, \citenamefont {Oberthaler}, \citenamefont
  {Schmied},\ and\ \citenamefont {Treutlein}}]{RMP2018:Smerzi}%
  \BibitemOpen
  \bibfield  {author} {\bibinfo {author} {\bibfnamefont {L.}~\bibnamefont
  {Pezz\`e}}, \bibinfo {author} {\bibfnamefont {A.}~\bibnamefont {Smerzi}},
  \bibinfo {author} {\bibfnamefont {M.~K.}\ \bibnamefont {Oberthaler}},
  \bibinfo {author} {\bibfnamefont {R.}~\bibnamefont {Schmied}}, \ and\
  \bibinfo {author} {\bibfnamefont {P.}~\bibnamefont {Treutlein}},\ }\bibfield
  {title} {\enquote {\bibinfo {title} {{Quantum metrology with nonclassical
  states of atomic ensembles}},}\ }\href {\doibase
  10.1103/RevModPhys.90.035005} {\bibfield  {journal} {\bibinfo  {journal}
  {Rev. Mod. Phys.}\ }\textbf {\bibinfo {volume} {90}},\ \bibinfo {pages}
  {035005} (\bibinfo {year} {2018})}\BibitemShut {NoStop}%
\bibitem [{\citenamefont {Braun}\ \emph {et~al.}(2018)\citenamefont {Braun},
  \citenamefont {Adesso}, \citenamefont {Benatti}, \citenamefont {Floreanini},
  \citenamefont {Marzolino}, \citenamefont {Mitchell},\ and\ \citenamefont
  {Pirandola}}]{RMP2018_HL:Braun}%
  \BibitemOpen
  \bibfield  {author} {\bibinfo {author} {\bibfnamefont {D.}~\bibnamefont
  {Braun}}, \bibinfo {author} {\bibfnamefont {G.}~\bibnamefont {Adesso}},
  \bibinfo {author} {\bibfnamefont {F.}~\bibnamefont {Benatti}}, \bibinfo
  {author} {\bibfnamefont {R.}~\bibnamefont {Floreanini}}, \bibinfo {author}
  {\bibfnamefont {U.}~\bibnamefont {Marzolino}}, \bibinfo {author}
  {\bibfnamefont {M.~W.}\ \bibnamefont {Mitchell}}, \ and\ \bibinfo {author}
  {\bibfnamefont {S.}~\bibnamefont {Pirandola}},\ }\bibfield  {title} {\enquote
  {\bibinfo {title} {{Quantum-enhanced measurements without entanglement}},}\
  }\href {\doibase 10.1103/RevModPhys.90.035006} {\bibfield  {journal}
  {\bibinfo  {journal} {Rev. Mod. Phys.}\ }\textbf {\bibinfo {volume} {90}},\
  \bibinfo {pages} {035006} (\bibinfo {year} {2018})}\BibitemShut {NoStop}%
\bibitem [{\citenamefont {Caves}\ \emph {et~al.}(1980)\citenamefont {Caves},
  \citenamefont {Thorne}, \citenamefont {Drever}, \citenamefont {Sandberg},\
  and\ \citenamefont {Zimmermann}}]{RevModPhys.52.341}%
  \BibitemOpen
  \bibfield  {author} {\bibinfo {author} {\bibfnamefont {C.~M.}\ \bibnamefont
  {Caves}}, \bibinfo {author} {\bibfnamefont {K.~S.}\ \bibnamefont {Thorne}},
  \bibinfo {author} {\bibfnamefont {R.~W.~P.}\ \bibnamefont {Drever}}, \bibinfo
  {author} {\bibfnamefont {V.~D.}\ \bibnamefont {Sandberg}}, \ and\ \bibinfo
  {author} {\bibfnamefont {M.}~\bibnamefont {Zimmermann}},\ }\bibfield  {title}
  {\enquote {\bibinfo {title} {{On the measurement of a weak classical force
  coupled to a quantum-mechanical oscillator. I. Issues of principle}},}\
  }\href {\doibase 10.1103/RevModPhys.52.341} {\bibfield  {journal} {\bibinfo
  {journal} {Rev. Mod. Phys.}\ }\textbf {\bibinfo {volume} {52}},\ \bibinfo
  {pages} {341--392} (\bibinfo {year} {1980})}\BibitemShut {NoStop}%
\bibitem [{\citenamefont {Caves}(1981)}]{PRD1981_SQL:Caves}%
  \BibitemOpen
  \bibfield  {author} {\bibinfo {author} {\bibfnamefont {C.~M.}\ \bibnamefont
  {Caves}},\ }\bibfield  {title} {\enquote {\bibinfo {title}
  {{Quantum-mechanical noise in an interferometer}},}\ }\href {\doibase
  10.1103/PhysRevD.23.1693} {\bibfield  {journal} {\bibinfo  {journal} {Phys.
  Rev. D}\ }\textbf {\bibinfo {volume} {23}},\ \bibinfo {pages} {1693--1708}
  (\bibinfo {year} {1981})}\BibitemShut {NoStop}%
\bibitem [{\citenamefont {Pitkin}\ \emph {et~al.}(2011)\citenamefont {Pitkin},
  \citenamefont {Reid}, \citenamefont {Rowan},\ and\ \citenamefont
  {Hough}}]{Pitkin2011}%
  \BibitemOpen
  \bibfield  {author} {\bibinfo {author} {\bibfnamefont {M.}~\bibnamefont
  {Pitkin}}, \bibinfo {author} {\bibfnamefont {S.}~\bibnamefont {Reid}},
  \bibinfo {author} {\bibfnamefont {S.}~\bibnamefont {Rowan}}, \ and\ \bibinfo
  {author} {\bibfnamefont {J.}~\bibnamefont {Hough}},\ }\bibfield  {title}
  {\enquote {\bibinfo {title} {{Gravitational Wave Detection by Interferometry
  (Ground and Space)}},}\ }\href {\doibase 10.12942/lrr-2011-5} {\bibfield
  {journal} {\bibinfo  {journal} {Living Reviews in Relativity}\ }\textbf
  {\bibinfo {volume} {14}},\ \bibinfo {pages} {5} (\bibinfo {year}
  {2011})}\BibitemShut {NoStop}%
\bibitem [{\citenamefont {Degen}\ \emph {et~al.}(2017)\citenamefont {Degen},
  \citenamefont {Reinhard},\ and\ \citenamefont
  {Cappellaro}}]{RevModPhys.89.035002}%
  \BibitemOpen
  \bibfield  {author} {\bibinfo {author} {\bibfnamefont {C.~L.}\ \bibnamefont
  {Degen}}, \bibinfo {author} {\bibfnamefont {F.}~\bibnamefont {Reinhard}}, \
  and\ \bibinfo {author} {\bibfnamefont {P.}~\bibnamefont {Cappellaro}},\
  }\bibfield  {title} {\enquote {\bibinfo {title} {{Quantum sensing}},}\ }\href
  {\doibase 10.1103/RevModPhys.89.035002} {\bibfield  {journal} {\bibinfo
  {journal} {Rev. Mod. Phys.}\ }\textbf {\bibinfo {volume} {89}},\ \bibinfo
  {pages} {035002} (\bibinfo {year} {2017})}\BibitemShut {NoStop}%
\bibitem [{\citenamefont {Wasilewski}\ \emph {et~al.}(2010)\citenamefont
  {Wasilewski}, \citenamefont {Jensen}, \citenamefont {Krauter}, \citenamefont
  {Renema}, \citenamefont {Balabas},\ and\ \citenamefont
  {Polzik}}]{PhysRevLett.104.133601}%
  \BibitemOpen
  \bibfield  {author} {\bibinfo {author} {\bibfnamefont {W.}~\bibnamefont
  {Wasilewski}}, \bibinfo {author} {\bibfnamefont {K.}~\bibnamefont {Jensen}},
  \bibinfo {author} {\bibfnamefont {H.}~\bibnamefont {Krauter}}, \bibinfo
  {author} {\bibfnamefont {J.~J.}\ \bibnamefont {Renema}}, \bibinfo {author}
  {\bibfnamefont {M.~V.}\ \bibnamefont {Balabas}}, \ and\ \bibinfo {author}
  {\bibfnamefont {E.~S.}\ \bibnamefont {Polzik}},\ }\bibfield  {title}
  {\enquote {\bibinfo {title} {{Quantum Noise Limited and Entanglement-Assisted
  Magnetometry}},}\ }\href {\doibase 10.1103/PhysRevLett.104.133601} {\bibfield
   {journal} {\bibinfo  {journal} {Phys. Rev. Lett.}\ }\textbf {\bibinfo
  {volume} {104}},\ \bibinfo {pages} {133601} (\bibinfo {year}
  {2010})}\BibitemShut {NoStop}%
\bibitem [{\citenamefont {Brask}\ \emph {et~al.}(2015)\citenamefont {Brask},
  \citenamefont {Chaves},\ and\ \citenamefont
  {Ko{\l}ody\'nski}}]{PhysRevX.5.031010}%
  \BibitemOpen
  \bibfield  {author} {\bibinfo {author} {\bibfnamefont {J.~B.}\ \bibnamefont
  {Brask}}, \bibinfo {author} {\bibfnamefont {R.}~\bibnamefont {Chaves}}, \
  and\ \bibinfo {author} {\bibfnamefont {J.}~\bibnamefont {Ko{\l}ody\'nski}},\
  }\bibfield  {title} {\enquote {\bibinfo {title} {{Improved Quantum
  Magnetometry beyond the Standard Quantum Limit}},}\ }\href {\doibase
  10.1103/PhysRevX.5.031010} {\bibfield  {journal} {\bibinfo  {journal} {Phys.
  Rev. X}\ }\textbf {\bibinfo {volume} {5}},\ \bibinfo {pages} {031010}
  (\bibinfo {year} {2015})}\BibitemShut {NoStop}%
\bibitem [{\citenamefont {Freier}\ \emph {et~al.}(2016)\citenamefont {Freier},
  \citenamefont {Hauth}, \citenamefont {Schkolnik}, \citenamefont {Leykauf},
  \citenamefont {Schilling}, \citenamefont {Wziontek}, \citenamefont
  {Scherneck}, \citenamefont {M\"uller},\ and\ \citenamefont
  {Peters}}]{Freier_2016}%
  \BibitemOpen
  \bibfield  {author} {\bibinfo {author} {\bibfnamefont {C.}~\bibnamefont
  {Freier}}, \bibinfo {author} {\bibfnamefont {M.}~\bibnamefont {Hauth}},
  \bibinfo {author} {\bibfnamefont {V.}~\bibnamefont {Schkolnik}}, \bibinfo
  {author} {\bibfnamefont {B.}~\bibnamefont {Leykauf}}, \bibinfo {author}
  {\bibfnamefont {M.}~\bibnamefont {Schilling}}, \bibinfo {author}
  {\bibfnamefont {H.}~\bibnamefont {Wziontek}}, \bibinfo {author}
  {\bibfnamefont {H.-G.}\ \bibnamefont {Scherneck}}, \bibinfo {author}
  {\bibfnamefont {J.}~\bibnamefont {M\"uller}}, \ and\ \bibinfo {author}
  {\bibfnamefont {A.}~\bibnamefont {Peters}},\ }\bibfield  {title} {\enquote
  {\bibinfo {title} {{Mobile quantum gravity sensor with unprecedented
  stability}},}\ }\href {\doibase 10.1088/1742-6596/723/1/012050} {\bibfield
  {journal} {\bibinfo  {journal} {Journal of Physics: Conference Series}\
  }\textbf {\bibinfo {volume} {723}},\ \bibinfo {pages} {012050} (\bibinfo
  {year} {2016})}\BibitemShut {NoStop}%
\bibitem [{\citenamefont {Fan}\ \emph {et~al.}(2015)\citenamefont {Fan},
  \citenamefont {Kumar}, \citenamefont {Sedlacek}, \citenamefont {K\"ubler},
  \citenamefont {Karimkashi},\ and\ \citenamefont {Shaffer}}]{Fan_2015}%
  \BibitemOpen
  \bibfield  {author} {\bibinfo {author} {\bibfnamefont {H.}~\bibnamefont
  {Fan}}, \bibinfo {author} {\bibfnamefont {S.}~\bibnamefont {Kumar}}, \bibinfo
  {author} {\bibfnamefont {J.}~\bibnamefont {Sedlacek}}, \bibinfo {author}
  {\bibfnamefont {H.}~\bibnamefont {K\"ubler}}, \bibinfo {author}
  {\bibfnamefont {S.}~\bibnamefont {Karimkashi}}, \ and\ \bibinfo {author}
  {\bibfnamefont {J.~P.}\ \bibnamefont {Shaffer}},\ }\bibfield  {title}
  {\enquote {\bibinfo {title} {{Atom based {RF} electric field sensing}},}\
  }\href {\doibase 10.1088/0953-4075/48/20/202001} {\bibfield  {journal}
  {\bibinfo  {journal} {J. Phys. B}\ }\textbf {\bibinfo {volume} {48}},\ \bibinfo {pages} {202001}
  (\bibinfo {year} {2015})}\BibitemShut {NoStop}%
\bibitem [{\citenamefont {{Dowling}}\ and\ \citenamefont
  {{Seshadreesan}}(2015)}]{6999929}%
  \BibitemOpen
  \bibfield  {author} {\bibinfo {author} {\bibfnamefont {J.~P.}\ \bibnamefont
  {{Dowling}}}\ and\ \bibinfo {author} {\bibfnamefont {K.~P.}\ \bibnamefont
  {{Seshadreesan}}},\ }\bibfield  {title} {\enquote {\bibinfo {title} {{Quantum
  Optical Technologies for Metrology, Sensing, and Imaging}},}\ }\href
  {\doibase 10.1109/JLT.2014.2386795} {\bibfield  {journal} {\bibinfo
  {journal} {Journal of Lightwave Technology}\ }\textbf {\bibinfo {volume}
  {33}},\ \bibinfo {pages} {2359--2370} (\bibinfo {year} {2015})}\BibitemShut
  {NoStop}%
\bibitem [{\citenamefont {{Thiebaut}}\ and\ \citenamefont
  {{Giovannelli}}(2010)}]{5355500}%
  \BibitemOpen
  \bibfield  {author} {\bibinfo {author} {\bibfnamefont {E.}~\bibnamefont
  {{Thiebaut}}}\ and\ \bibinfo {author} {\bibfnamefont {J.}~\bibnamefont
  {{Giovannelli}}},\ }\bibfield  {title} {\enquote {\bibinfo {title} {{Image
  reconstruction in optical interferometry}},}\ }\href {\doibase
  10.1109/MSP.2009.934870} {\bibfield  {journal} {\bibinfo  {journal} {IEEE
  Signal Processing Magazine}\ }\textbf {\bibinfo {volume} {27}},\ \bibinfo
  {pages} {97--109} (\bibinfo {year} {2010})}\BibitemShut {NoStop}%
\bibitem [{\citenamefont {Brida}\ \emph {et~al.}(2010)\citenamefont {Brida},
  \citenamefont {Genovese},\ and\ \citenamefont
  {Berchera}}]{NaturePhotonicsvolume4pages227to230_2010}%
  \BibitemOpen
  \bibfield  {author} {\bibinfo {author} {\bibfnamefont {G.}~\bibnamefont
  {Brida}}, \bibinfo {author} {\bibfnamefont {M.}~\bibnamefont {Genovese}}, \
  and\ \bibinfo {author} {\bibfnamefont {I.~R.}\ \bibnamefont {Berchera}},\
  }\bibfield  {title} {\enquote {\bibinfo {title} {{Experimental realization of
  sub-shot-noise quantum imaging}},}\ }\href@noop {} {\bibfield  {journal}
  {\bibinfo  {journal} {Nat. Photonics}\ }\textbf {\bibinfo {volume} {4}},\
  \bibinfo {pages} {227} (\bibinfo {year} {2010})}\BibitemShut {NoStop}%
\bibitem [{\citenamefont {Yurke}\ \emph {et~al.}(1986)\citenamefont {Yurke},
  \citenamefont {McCall},\ and\ \citenamefont {Klauder}}]{PRA1986_SU11:Yurke}%
  \BibitemOpen
  \bibfield  {author} {\bibinfo {author} {\bibfnamefont {B.}~\bibnamefont
  {Yurke}}, \bibinfo {author} {\bibfnamefont {S.~L.}\ \bibnamefont {McCall}}, \
  and\ \bibinfo {author} {\bibfnamefont {J.~R.}\ \bibnamefont {Klauder}},\
  }\bibfield  {title} {\enquote {\bibinfo {title} {{SU(2) and SU(1,1)
  interferometers}},}\ }\href {\doibase 10.1103/PhysRevA.33.4033} {\bibfield
  {journal} {\bibinfo  {journal} {Phys. Rev. A}\ }\textbf {\bibinfo {volume}
  {33}},\ \bibinfo {pages} {4033--4054} (\bibinfo {year} {1986})}\BibitemShut
  {NoStop}%
\bibitem [{\citenamefont {Xiao}\ \emph {et~al.}(1987)\citenamefont {Xiao},
  \citenamefont {Wu},\ and\ \citenamefont {Kimble}}]{PRL1987_SNL:Xiao}%
  \BibitemOpen
  \bibfield  {author} {\bibinfo {author} {\bibfnamefont {M.}~\bibnamefont
  {Xiao}}, \bibinfo {author} {\bibfnamefont {L.-A.}\ \bibnamefont {Wu}}, \ and\
  \bibinfo {author} {\bibfnamefont {H.~J.}\ \bibnamefont {Kimble}},\ }\bibfield
   {title} {\enquote {\bibinfo {title} {{Precision measurement beyond the
  shot-noise limit}},}\ }\href {\doibase 10.1103/PhysRevLett.59.278} {\bibfield
   {journal} {\bibinfo  {journal} {Phys. Rev. Lett.}\ }\textbf {\bibinfo
  {volume} {59}},\ \bibinfo {pages} {278--281} (\bibinfo {year}
  {1987})}\BibitemShut {NoStop}%
\bibitem [{\citenamefont {Holland}\ and\ \citenamefont
  {Burnett}(1993)}]{PRL1993_HL:Burnett}%
  \BibitemOpen
  \bibfield  {author} {\bibinfo {author} {\bibfnamefont {M.~J.}\ \bibnamefont
  {Holland}}\ and\ \bibinfo {author} {\bibfnamefont {K.}~\bibnamefont
  {Burnett}},\ }\bibfield  {title} {\enquote {\bibinfo {title}
  {{Interferometric detection of optical phase shifts at the Heisenberg
  limit}},}\ }\href {\doibase 10.1103/PhysRevLett.71.1355} {\bibfield
  {journal} {\bibinfo  {journal} {Phys. Rev. Lett.}\ }\textbf {\bibinfo
  {volume} {71}},\ \bibinfo {pages} {1355--1358} (\bibinfo {year}
  {1993})}\BibitemShut {NoStop}%
\bibitem [{\citenamefont {Braunstein}\ and\ \citenamefont
  {Caves}(1994)}]{PRL1994_QFI:Braunstein}%
  \BibitemOpen
  \bibfield  {author} {\bibinfo {author} {\bibfnamefont {S.~L.}\ \bibnamefont
  {Braunstein}}\ and\ \bibinfo {author} {\bibfnamefont {C.~M.}\ \bibnamefont
  {Caves}},\ }\bibfield  {title} {\enquote {\bibinfo {title} {{Statistical
  distance and the geometry of quantum states}},}\ }\href {\doibase
  10.1103/PhysRevLett.72.3439} {\bibfield  {journal} {\bibinfo  {journal}
  {Phys. Rev. Lett.}\ }\textbf {\bibinfo {volume} {72}},\ \bibinfo {pages}
  {3439--3443} (\bibinfo {year} {1994})}\BibitemShut {NoStop}%
\bibitem [{\citenamefont {Pezz\'e}\ and\ \citenamefont
  {Smerzi}(2009)}]{PRL2009_Hl:Pezz}%
  \BibitemOpen
  \bibfield  {author} {\bibinfo {author} {\bibfnamefont {L.}~\bibnamefont
  {Pezz\'e}}\ and\ \bibinfo {author} {\bibfnamefont {A.}~\bibnamefont
  {Smerzi}},\ }\bibfield  {title} {\enquote {\bibinfo {title} {{Entanglement,
  Nonlinear Dynamics, and the Heisenberg Limit}},}\ }\href {\doibase
  10.1103/PhysRevLett.102.100401} {\bibfield  {journal} {\bibinfo  {journal}
  {Phys. Rev. Lett.}\ }\textbf {\bibinfo {volume} {102}},\ \bibinfo {pages}
  {100401} (\bibinfo {year} {2009})}\BibitemShut {NoStop}%
\bibitem [{\citenamefont {Boixo}\ \emph {et~al.}(2009)\citenamefont {Boixo},
  \citenamefont {Datta}, \citenamefont {Davis}, \citenamefont {Shaji},
  \citenamefont {Tacla},\ and\ \citenamefont {Caves}}]{PRA2009_HL:Boixo}%
  \BibitemOpen
  \bibfield  {author} {\bibinfo {author} {\bibfnamefont {S.}~\bibnamefont
  {Boixo}}, \bibinfo {author} {\bibfnamefont {A.}~\bibnamefont {Datta}},
  \bibinfo {author} {\bibfnamefont {M.~J.}\ \bibnamefont {Davis}}, \bibinfo
  {author} {\bibfnamefont {A.}~\bibnamefont {Shaji}}, \bibinfo {author}
  {\bibfnamefont {A.~B.}\ \bibnamefont {Tacla}}, \ and\ \bibinfo {author}
  {\bibfnamefont {C.~M.}\ \bibnamefont {Caves}},\ }\bibfield  {title} {\enquote
  {\bibinfo {title} {{Quantum-limited metrology and Bose-Einstein
  condensates}},}\ }\href {\doibase 10.1103/PhysRevA.80.032103} {\bibfield
  {journal} {\bibinfo  {journal} {Phys. Rev. A}\ }\textbf {\bibinfo {volume}
  {80}},\ \bibinfo {pages} {032103} (\bibinfo {year} {2009})}\BibitemShut
  {NoStop}%
\bibitem [{\citenamefont {Kawaguchi}\ and\ \citenamefont
  {Ueda}(2012)}]{PR2012_Spinor_BEC:Ueda}%
  \BibitemOpen
  \bibfield  {author} {\bibinfo {author} {\bibfnamefont {Y.}~\bibnamefont
  {Kawaguchi}}\ and\ \bibinfo {author} {\bibfnamefont {M.}~\bibnamefont
  {Ueda}},\ }\bibfield  {title} {\enquote {\bibinfo {title} {{Spinor
  Bose–Einstein condensates}},}\ }\href {\doibase
  https://doi.org/10.1016/j.physrep.2012.07.005} {\bibfield  {journal}
  {\bibinfo  {journal} {Physics Reports}\ }\textbf {\bibinfo {volume} {520}},\
  \bibinfo {pages} {253 -- 381} (\bibinfo {year} {2012})}\BibitemShut {NoStop}%
\bibitem [{\citenamefont {Stamper-Kurn}\ and\ \citenamefont
  {Ueda}(2013)}]{RMP2013_Spinor_BEC:Ueda}%
  \BibitemOpen
  \bibfield  {author} {\bibinfo {author} {\bibfnamefont {D.~M.}\ \bibnamefont
  {Stamper-Kurn}}\ and\ \bibinfo {author} {\bibfnamefont {M.}~\bibnamefont
  {Ueda}},\ }\bibfield  {title} {\enquote {\bibinfo {title} {{Spinor Bose
  gases: Symmetries, magnetism, and quantum dynamics}},}\ }\href {\doibase
  10.1103/RevModPhys.85.1191} {\bibfield  {journal} {\bibinfo  {journal} {Rev.
  Mod. Phys.}\ }\textbf {\bibinfo {volume} {85}},\ \bibinfo {pages}
  {1191--1244} (\bibinfo {year} {2013})}\BibitemShut {NoStop}%
\bibitem [{\citenamefont {Yi}\ \emph {et~al.}(2002)\citenamefont {Yi},
  \citenamefont {M\"ustecapl\ifmmode \imath \else \i
  \fi{}o\ifmmode~\breve{g}\else \u{g}\fi{}lu}, \citenamefont {Sun},\ and\
  \citenamefont {You}}]{PRA2002_SMA:Yi}%
  \BibitemOpen
  \bibfield  {author} {\bibinfo {author} {\bibfnamefont {S.}~\bibnamefont
  {Yi}}, \bibinfo {author} {\bibfnamefont {\"O.~E.}\ \bibnamefont
  {M\"ustecapl\ifmmode \imath \else \i \fi{}o\ifmmode~\breve{g}\else
  \u{g}\fi{}lu}}, \bibinfo {author} {\bibfnamefont {C.~P.}\ \bibnamefont
  {Sun}}, \ and\ \bibinfo {author} {\bibfnamefont {L.}~\bibnamefont {You}},\
  }\bibfield  {title} {\enquote {\bibinfo {title} {{Single-mode approximation
  in a spinor-1 atomic condensate}},}\ }\href {\doibase
  10.1103/PhysRevA.66.011601} {\bibfield  {journal} {\bibinfo  {journal} {Phys.
  Rev. A}\ }\textbf {\bibinfo {volume} {66}},\ \bibinfo {pages} {011601}
  (\bibinfo {year} {2002})}\BibitemShut {NoStop}%
\bibitem [{\citenamefont {Gabbrielli}\ \emph {et~al.}(2015)\citenamefont
  {Gabbrielli}, \citenamefont {Pezz\`e},\ and\ \citenamefont
  {Smerzi}}]{PRL2015_SU11:Gabbrielli}%
  \BibitemOpen
  \bibfield  {author} {\bibinfo {author} {\bibfnamefont {M.}~\bibnamefont
  {Gabbrielli}}, \bibinfo {author} {\bibfnamefont {L.}~\bibnamefont {Pezz\`e}},
  \ and\ \bibinfo {author} {\bibfnamefont {A.}~\bibnamefont {Smerzi}},\
  }\bibfield  {title} {\enquote {\bibinfo {title} {{Spin-Mixing Interferometry
  with Bose-Einstein Condensates}},}\ }\href {\doibase
  10.1103/PhysRevLett.115.163002} {\bibfield  {journal} {\bibinfo  {journal}
  {Phys. Rev. Lett.}\ }\textbf {\bibinfo {volume} {115}},\ \bibinfo {pages}
  {163002} (\bibinfo {year} {2015})}\BibitemShut {NoStop}%
\bibitem [{\citenamefont {Linnemann}\ \emph {et~al.}(2016)\citenamefont
  {Linnemann}, \citenamefont {Strobel}, \citenamefont {Muessel}, \citenamefont
  {Schulz}, \citenamefont {Lewis-Swan}, \citenamefont {Kheruntsyan},\ and\
  \citenamefont {Oberthaler}}]{PRL2016_SU11:Linnemann}%
  \BibitemOpen
  \bibfield  {author} {\bibinfo {author} {\bibfnamefont {D.}~\bibnamefont
  {Linnemann}}, \bibinfo {author} {\bibfnamefont {H.}~\bibnamefont {Strobel}},
  \bibinfo {author} {\bibfnamefont {W.}~\bibnamefont {Muessel}}, \bibinfo
  {author} {\bibfnamefont {J.}~\bibnamefont {Schulz}}, \bibinfo {author}
  {\bibfnamefont {R.~J.}\ \bibnamefont {Lewis-Swan}}, \bibinfo {author}
  {\bibfnamefont {K.~V.}\ \bibnamefont {Kheruntsyan}}, \ and\ \bibinfo {author}
  {\bibfnamefont {M.~K.}\ \bibnamefont {Oberthaler}},\ }\bibfield  {title}
  {\enquote {\bibinfo {title} {{Quantum-Enhanced Sensing Based on Time Reversal
  of Nonlinear Dynamics}},}\ }\href {\doibase 10.1103/PhysRevLett.117.013001}
  {\bibfield  {journal} {\bibinfo  {journal} {Phys. Rev. Lett.}\ }\textbf
  {\bibinfo {volume} {117}},\ \bibinfo {pages} {013001} (\bibinfo {year}
  {2016})}\BibitemShut {NoStop}%
\bibitem [{\citenamefont {Law}\ \emph {et~al.}(1998)\citenamefont {Law},
  \citenamefont {Pu},\ and\ \citenamefont {Bigelow}}]{PRL1998_SMD:Law}%
  \BibitemOpen
  \bibfield  {author} {\bibinfo {author} {\bibfnamefont {C.~K.}\ \bibnamefont
  {Law}}, \bibinfo {author} {\bibfnamefont {H.}~\bibnamefont {Pu}}, \ and\
  \bibinfo {author} {\bibfnamefont {N.~P.}\ \bibnamefont {Bigelow}},\
  }\bibfield  {title} {\enquote {\bibinfo {title} {{Quantum Spins Mixing in
  Spinor Bose-Einstein Condensates}},}\ }\href {\doibase
  10.1103/PhysRevLett.81.5257} {\bibfield  {journal} {\bibinfo  {journal}
  {Phys. Rev. Lett.}\ }\textbf {\bibinfo {volume} {81}},\ \bibinfo {pages}
  {5257--5261} (\bibinfo {year} {1998})}\BibitemShut {NoStop}%
\bibitem [{\citenamefont {Gerbier}\ \emph {et~al.}(2006)\citenamefont
  {Gerbier}, \citenamefont {Widera}, \citenamefont {F\"olling}, \citenamefont
  {Mandel},\ and\ \citenamefont {Bloch}}]{PRA2006_MW_dressing:Gerbier}%
  \BibitemOpen
  \bibfield  {author} {\bibinfo {author} {\bibfnamefont {F.}~\bibnamefont
  {Gerbier}}, \bibinfo {author} {\bibfnamefont {A.}~\bibnamefont {Widera}},
  \bibinfo {author} {\bibfnamefont {S.}~\bibnamefont {F\"olling}}, \bibinfo
  {author} {\bibfnamefont {O.}~\bibnamefont {Mandel}}, \ and\ \bibinfo {author}
  {\bibfnamefont {I.}~\bibnamefont {Bloch}},\ }\bibfield  {title} {\enquote
  {\bibinfo {title} {{Resonant control of spin dynamics in ultracold quantum
  gases by microwave dressing}},}\ }\href {\doibase 10.1103/PhysRevA.73.041602}
  {\bibfield  {journal} {\bibinfo  {journal} {Phys. Rev. A}\ }\textbf {\bibinfo
  {volume} {73}},\ \bibinfo {pages} {041602} (\bibinfo {year}
  {2006})}\BibitemShut {NoStop}%
\bibitem [{\citenamefont {Zhao}\ \emph {et~al.}(2014)\citenamefont {Zhao},
  \citenamefont {Jiang}, \citenamefont {Tang}, \citenamefont {Webb},\ and\
  \citenamefont {Liu}}]{PRA2014_MW_dressing:Zhao}%
  \BibitemOpen
  \bibfield  {author} {\bibinfo {author} {\bibfnamefont {L.}~\bibnamefont
  {Zhao}}, \bibinfo {author} {\bibfnamefont {J.}~\bibnamefont {Jiang}},
  \bibinfo {author} {\bibfnamefont {T.}~\bibnamefont {Tang}}, \bibinfo {author}
  {\bibfnamefont {M.}~\bibnamefont {Webb}}, \ and\ \bibinfo {author}
  {\bibfnamefont {Y.}~\bibnamefont {Liu}},\ }\bibfield  {title} {\enquote
  {\bibinfo {title} {{Dynamics in spinor condensates tuned by a microwave
  dressing field}},}\ }\href {\doibase 10.1103/PhysRevA.89.023608} {\bibfield
  {journal} {\bibinfo  {journal} {Phys. Rev. A}\ }\textbf {\bibinfo {volume}
  {89}},\ \bibinfo {pages} {023608} (\bibinfo {year} {2014})}\BibitemShut
  {NoStop}%
\bibitem [{\citenamefont {Chin}\ \emph {et~al.}(2010)\citenamefont {Chin},
  \citenamefont {Grimm}, \citenamefont {Julienne},\ and\ \citenamefont
  {Tiesinga}}]{RMP2010_FR:Chin}%
  \BibitemOpen
  \bibfield  {author} {\bibinfo {author} {\bibfnamefont {C.}~\bibnamefont
  {Chin}}, \bibinfo {author} {\bibfnamefont {R.}~\bibnamefont {Grimm}},
  \bibinfo {author} {\bibfnamefont {P.}~\bibnamefont {Julienne}}, \ and\
  \bibinfo {author} {\bibfnamefont {E.}~\bibnamefont {Tiesinga}},\ }\bibfield
  {title} {\enquote {\bibinfo {title} {{Feshbach resonances in ultracold
  gases}},}\ }\href {\doibase 10.1103/RevModPhys.82.1225} {\bibfield  {journal}
  {\bibinfo  {journal} {Rev. Mod. Phys.}\ }\textbf {\bibinfo {volume} {82}},\
  \bibinfo {pages} {1225--1286} (\bibinfo {year} {2010})}\BibitemShut {NoStop}%
\bibitem [{\citenamefont {Choi}\ and\ \citenamefont
  {Sundaram}(2008)}]{PRA2008_Hl:Choi}%
  \BibitemOpen
  \bibfield  {author} {\bibinfo {author} {\bibfnamefont {S.}~\bibnamefont
  {Choi}}\ and\ \bibinfo {author} {\bibfnamefont {B.}~\bibnamefont
  {Sundaram}},\ }\bibfield  {title} {\enquote {\bibinfo {title} {{Bose-Einstein
  condensate as a nonlinear Ramsey interferometer operating beyond the
  Heisenberg limit}},}\ }\href {\doibase 10.1103/PhysRevA.77.053613} {\bibfield
   {journal} {\bibinfo  {journal} {Phys. Rev. A}\ }\textbf {\bibinfo {volume}
  {77}},\ \bibinfo {pages} {053613} (\bibinfo {year} {2008})}\BibitemShut
  {NoStop}%
\bibitem [{\citenamefont {Napolitano}\ \emph {et~al.}(2011)\citenamefont
  {Napolitano}, \citenamefont {Koschorreck}, \citenamefont {Dubost},
  \citenamefont {Behbood}, \citenamefont {Sewell},\ and\ \citenamefont
  {Mitchell}}]{Nature2011_HL:Napolitano}%
  \BibitemOpen
  \bibfield  {author} {\bibinfo {author} {\bibfnamefont {M.}~\bibnamefont
  {Napolitano}}, \bibinfo {author} {\bibfnamefont {M.}~\bibnamefont
  {Koschorreck}}, \bibinfo {author} {\bibfnamefont {B.}~\bibnamefont {Dubost}},
  \bibinfo {author} {\bibfnamefont {N.}~\bibnamefont {Behbood}}, \bibinfo
  {author} {\bibfnamefont {R.~J.}\ \bibnamefont {Sewell}}, \ and\ \bibinfo
  {author} {\bibfnamefont {M.~W.}\ \bibnamefont {Mitchell}},\ }\bibfield
  {title} {\enquote {\bibinfo {title} {{Interaction-based quantum metrology
  showing scaling beyond the Heisenberg limit}},}\ }\href {\doibase
  10.1038/nature09778} {\bibfield  {journal} {\bibinfo  {journal} {Nature}\
  }\textbf {\bibinfo {volume} {471}},\ \bibinfo {pages} {486} (\bibinfo {year}
  {2011})}\BibitemShut {NoStop}%
\bibitem [{\citenamefont {Anisimov}\ \emph {et~al.}(2010)\citenamefont
  {Anisimov}, \citenamefont {Raterman}, \citenamefont {Chiruvelli},
  \citenamefont {Plick}, \citenamefont {Huver}, \citenamefont {Lee},\ and\
  \citenamefont {Dowling}}]{PRL2010_HL:Anisimov}%
  \BibitemOpen
  \bibfield  {author} {\bibinfo {author} {\bibfnamefont {P.~M.}\ \bibnamefont
  {Anisimov}}, \bibinfo {author} {\bibfnamefont {G.~M.}\ \bibnamefont
  {Raterman}}, \bibinfo {author} {\bibfnamefont {A.}~\bibnamefont
  {Chiruvelli}}, \bibinfo {author} {\bibfnamefont {W.~N.}\ \bibnamefont
  {Plick}}, \bibinfo {author} {\bibfnamefont {S.~D.}\ \bibnamefont {Huver}},
  \bibinfo {author} {\bibfnamefont {H.}~\bibnamefont {Lee}}, \ and\ \bibinfo
  {author} {\bibfnamefont {J.~P.}\ \bibnamefont {Dowling}},\ }\bibfield
  {title} {\enquote {\bibinfo {title} {{Quantum Metrology with Two-Mode
  Squeezed Vacuum: Parity Detection Beats the Heisenberg Limit}},}\ }\href
  {\doibase 10.1103/PhysRevLett.104.103602} {\bibfield  {journal} {\bibinfo
  {journal} {Phys. Rev. Lett.}\ }\textbf {\bibinfo {volume} {104}},\ \bibinfo
  {pages} {103602} (\bibinfo {year} {2010})}\BibitemShut {NoStop}%
\bibitem [{\citenamefont {Tsarev}\ \emph {et~al.}(2018)\citenamefont {Tsarev},
  \citenamefont {Arakelian}, \citenamefont {Chuang}, \citenamefont {Lee},\ and\
  \citenamefont {Alodjants}}]{OSA2018_HL:Tsarev}%
  \BibitemOpen
  \bibfield  {author} {\bibinfo {author} {\bibfnamefont {D.~V.}\ \bibnamefont
  {Tsarev}}, \bibinfo {author} {\bibfnamefont {S.~M.}\ \bibnamefont
  {Arakelian}}, \bibinfo {author} {\bibfnamefont {Y.-L.}\ \bibnamefont
  {Chuang}}, \bibinfo {author} {\bibfnamefont {R.-K.}\ \bibnamefont {Lee}}, \
  and\ \bibinfo {author} {\bibfnamefont {A.~P.}\ \bibnamefont {Alodjants}},\
  }\bibfield  {title} {\enquote {\bibinfo {title} {Quantum metrology beyond
  heisenberg limit with entangled matter wave solitons},}\ }\href {\doibase
  10.1364/OE.26.019583} {\bibfield  {journal} {\bibinfo  {journal} {Opt.
  Express}\ }\textbf {\bibinfo {volume} {26}},\ \bibinfo {pages} {19583--19595}
  (\bibinfo {year} {2018})}\BibitemShut {NoStop}%
\bibitem [{\citenamefont {Szigeti}\ \emph {et~al.}(2017)\citenamefont
  {Szigeti}, \citenamefont {Lewis-Swan},\ and\ \citenamefont
  {Haine}}]{PRL2017_Pump_up_SU11:Szigeti}%
  \BibitemOpen
  \bibfield  {author} {\bibinfo {author} {\bibfnamefont {S.~S.}\ \bibnamefont
  {Szigeti}}, \bibinfo {author} {\bibfnamefont {R.~J.}\ \bibnamefont
  {Lewis-Swan}}, \ and\ \bibinfo {author} {\bibfnamefont {S.~A.}\ \bibnamefont
  {Haine}},\ }\bibfield  {title} {\enquote {\bibinfo {title} {{Pumped-Up
  SU(1,1) Interferometry}},}\ }\href {\doibase 10.1103/PhysRevLett.118.150401}
  {\bibfield  {journal} {\bibinfo  {journal} {Phys. Rev. Lett.}\ }\textbf
  {\bibinfo {volume} {118}},\ \bibinfo {pages} {150401} (\bibinfo {year}
  {2017})}\BibitemShut {NoStop}%
\bibitem [{\citenamefont {Hamley}\ \emph {et~al.}(2012)\citenamefont {Hamley},
  \citenamefont {Gerving}, \citenamefont {Hoang}, \citenamefont {Bookjans},\
  and\ \citenamefont {Chapman}}]{NP2012_Amplification:Hamley}%
  \BibitemOpen
  \bibfield  {author} {\bibinfo {author} {\bibfnamefont {C.~D.}\ \bibnamefont
  {Hamley}}, \bibinfo {author} {\bibfnamefont {C.~S.}\ \bibnamefont {Gerving}},
  \bibinfo {author} {\bibfnamefont {T.~M.}\ \bibnamefont {Hoang}}, \bibinfo
  {author} {\bibfnamefont {E.~M.}\ \bibnamefont {Bookjans}}, \ and\ \bibinfo
  {author} {\bibfnamefont {M.~S.}\ \bibnamefont {Chapman}},\ }\bibfield
  {title} {\enquote {\bibinfo {title} {{Spin-nematic squeezed vacuum in a
  quantum gas}},}\ }\href {\doibase 10.1038/nphys2245} {\bibfield  {journal}
  {\bibinfo  {journal} {Nature Physics}\ }\textbf {\bibinfo {volume} {8}},\
  \bibinfo {pages} {305} (\bibinfo {year} {2012})}\BibitemShut {NoStop}%
\bibitem [{\citenamefont {Wrubel}\ \emph {et~al.}(2018)\citenamefont {Wrubel},
  \citenamefont {Schwettmann}, \citenamefont {Fahey}, \citenamefont {Glassman},
  \citenamefont {Pechkis}, \citenamefont {Griffin}, \citenamefont {Barnett},
  \citenamefont {Tiesinga},\ and\ \citenamefont
  {Lett}}]{PRA2018_SU11:P.D.Lett}%
  \BibitemOpen
  \bibfield  {author} {\bibinfo {author} {\bibfnamefont {J.~P.}\ \bibnamefont
  {Wrubel}}, \bibinfo {author} {\bibfnamefont {A.}~\bibnamefont {Schwettmann}},
  \bibinfo {author} {\bibfnamefont {D.~P.}\ \bibnamefont {Fahey}}, \bibinfo
  {author} {\bibfnamefont {Z.}~\bibnamefont {Glassman}}, \bibinfo {author}
  {\bibfnamefont {H.~K.}\ \bibnamefont {Pechkis}}, \bibinfo {author}
  {\bibfnamefont {P.~F.}\ \bibnamefont {Griffin}}, \bibinfo {author}
  {\bibfnamefont {R.}~\bibnamefont {Barnett}}, \bibinfo {author} {\bibfnamefont
  {E.}~\bibnamefont {Tiesinga}}, \ and\ \bibinfo {author} {\bibfnamefont
  {P.~D.}\ \bibnamefont {Lett}},\ }\bibfield  {title} {\enquote {\bibinfo
  {title} {{Spinor Bose-Einstein-condensate phase-sensitive amplifier for
  SU(1,1) interferometry}},}\ }\href {\doibase 10.1103/PhysRevA.98.023620}
  {\bibfield  {journal} {\bibinfo  {journal} {Phys. Rev. A}\ }\textbf {\bibinfo
  {volume} {98}},\ \bibinfo {pages} {023620} (\bibinfo {year}
  {2018})}\BibitemShut {NoStop}%
\bibitem [{\citenamefont {Scully}\ and\ \citenamefont
  {Zubairy}(1997)}]{scully1999quantum}%
  \BibitemOpen
  \bibfield  {author} {\bibinfo {author} {\bibfnamefont {M.~O.}\ \bibnamefont
  {Scully}}\ and\ \bibinfo {author} {\bibfnamefont {M.~S.}\ \bibnamefont
  {Zubairy}},\ }\href@noop {} {\emph {\bibinfo {title} {{Quantum optics}}}}\
  (\bibinfo  {publisher} {Cambridge University Press},\ \bibinfo {year}
  {1997})\BibitemShut {NoStop}%
\bibitem {PhysRevA.83.042704}
  S. Knoop, T. Schuster, R. Scelle, A Trautmann, J. Appmeier, 
M. K. Oberthaler, E. Tiesinga, and E. Tiemann, 
``Feshbach spectroscopy and analysis of the interaction potentials of ultracold sodium,'' 
  Phys. Rev. A
  {\bf{83}},
   {042704},
  (2011).





\end{thebibliography}

\end{document}